\documentclass[MSNbibl,number,citesort,dvips]{arxstspdf}
\usepackage{flushend}
\usepackage{stfloats}

\volume{27}
\issue{2}
\pubyear{2012}
\firstpage{278}
\lastpage{293}
\doi{10.1214/11-STS371}

\makeatletter
\newtheorem{thmm}{Theorem}
\newproclaim{example}{Example}
\newproclaim{rem}{Remark}
\newtheorem{prop}{Proposition}
\newtheorem{lem}{Lemma}
\newproclaim{defn}{Definition}
\makeatother

\begin{document}
\begin{frontmatter}

\title{Estimation in Discrete Parameter Models}
\runtitle{Discrete Parameter Models}

\begin{aug}
\author[a]{\fnms{Christine} \snm{Choirat}\corref{}\ead[label=e1]{cchoirat@unav.es}}
\and
\author[b]{\fnms{Raffaello} \snm{Seri}\ead[label=e2]{raffaello.seri@uninsubria.it}}
\runauthor{C. Choirat and R. Seri}

\affiliation{Universidad de Navarra and Universit\`a degli Studi dell'Insubria}

\address[a]{Christine Choirat is Associate Professor, Department of Economics,
School of Economics and Business Management,
Universidad de Navarra,
Edificio de Bibliotecas (Entrada Este),
31080 Pamplona, Spain \printead{e1}.}
\address[b]{Raffaello Seri is Assistant Professor, Dipartimento di Economia,
Universit\`a degli Studi dell'Insubria,
Via Monte Generoso 71,
21100 Varese, Italy
\printead{e2}.}

\end{aug}

%
\begin{abstract}
In some estimation problems, especially in applications
dealing with information theory, signal processing and biology, theory
provides us with additional information allowing us to restrict the
parameter space to a finite number of points. In this case, we speak
of discrete parameter models. Even though the problem is quite old and
has interesting connections
with testing and model selection, asymptotic theory for these models
has hardly ever been studied. Therefore, we discuss consistency, asymptotic
distribution theory, information inequalities and their relations
with efficiency and superefficiency for a~general class of $m$-estimators.
\end{abstract}

%

\end{frontmatter}

\section{Introduction}

Sometimes, especially in applications dealing with signal processing
and biology, theory provides us with some additional information allowing
us to restrict the parameter space to a finite number of points; in
these cases, we speak of \textit{discrete parameter models}. Statistical
inference when the parameter space is reduced to a lattice was first
considered by Hammersley \cite{Hammersley-JRSSB-1950} in a seminal
paper. However, since the
author was motivated by the measurement of the mean weight of insulin,
he focused mainly on the case of a Gaussian distribution with known
variance and unknown integer mean (see \cite{Hammersley-JRSSB-1950},
page~192); this case was further developed by Khan \cite
{Khan-AoS-1973,Khan-CJS-1978,Khan-JSPI-2000,Khan-IJMMS-2003}.
The Poisson case also met some attention in the literature and was
dealt with by Hammersley (\cite{Hammersley-JRSSB-1950}, page~199) and
others \cite{McCabe-AoMS-1972,Stark-JASA-1975}.

Previous works have
shown that the rate of convergence of $m$-estimators is often exponential
[\cite*{Hammersley-JRSSB-1950,Vajda-CMJ-1967}, \cite*{Vajda-AM-1971a,Vajda-AM-1971b}].
General treatments of admissibility and related topics are in \cite
{Robson-AoMS-1958,Gersanov-Samroni-TPA-1976,Hsuan-AoS-1979,Meeden-Ghosh-AoS-1981}
(see also the book \cite{Berger-BOOK-1993}); special cases have been
dealt with in~\cite{Karlin-AoMS-1958} (page~424, for the case
of a translation integral parameter and of integral data under the
quadratic loss), \cite
{Hammersley-JRSSB-1950,Khan-AoS-1973,Ghosh-Meeden-San-1978,Khan-CJS-1978,Khan-JSPI-2000,Khan-IJMMS-2003}
(for the case of the Gaussian distribution) and \cite{Blyth-AoS-1974}
(for the case of the discrete uniform distribution). Other papers
dealing with optimality in discrete parameter spaces are \cite
{Vajda-Kyb-1967a,Vajda-Kyb-1967b,Vajda-BOOK-1968,Vajda-BOOK-1974,Gersanov-TPA-1979}.
Optimality of estimation under a discrete parameter space was also
considered by Vajda \cite{Vajda-CMJ-1967,Vajda-AM-1971a,Vajda-AM-1971b}
in a nonorthodox setting inspired by R{\'e}nyi's theory of random search.
Other aspects that have been studied are Bayesian encompassing \cite
{Florens-Richard-WP-1989},
construction of confidence intervals (\cite{Cox-Hinkley-BOOK-1974},
pages 224--225),
comparison of statistical experiments (\cite
{Torgersen-ZWvG-1970}, \cite{LeCam-Yang-BOOK-1990}, Section 2.2),
sufficiency and minimal sufficiency \cite{LaMotte-AS-2008} and
best prediction \cite{Teunissen-JoG-2007}. Moreover, in the estimation
of complex statistical models (see \cite
{Grenander-BOOK-1981}, \cite{Clement-Thesis-1995},
Chapter~4) and in the calculation of efficiency rates (see \cite
{Bahadur-San-1960,LeCam-Yang-BOOK-1990,Chamberlain-JoAE-2000}),
approximating a general parameter space by a sequence of finite sets
has proved to be a valuable tool. A~few papers showed the practical
importance of discrete parameter
models in signal processing, automatic control and information theory
and derived some bounds on the performance of the estimators (see
\cite{Lainiotis-IEEETIT-1969a,Lainiotis-IEEETIT-1969b,Liporace-IEEETIT-1971,Hawkes-Moore-IEEETAC-1976,Hawkes-Moore-IEEETIT-1976,Hawkes-Moore-PIEEE-1976,Baram-Sandell-BOOK-1977,Baram-IEEETAC-1978,Baram-Sandell-IEEETAC-1978a,Baram-Sandell-IEEETAC-1978b}).
More recently, the topic has received new interest in the information
theory literature (see \cite{Poor-Verdu-IEEETIT-1995,Kanaya-Han-IEEETIT-1995},
and the review paper \cite{Hero-HoSP-1999}), in stochastic integer
programming (see \cite
{Futschik-Pflug-AoOR-1995,Kleywegt-Shapiro-Homem-de-Mello-SIAMJO-2001,vanderVlerk-WWW-2003}),
and in geodesy (see, e.g.,~\cite{Teunissen-JoG-2007}, Section 5).

However, no general formula for the convergence rate has ever been
obtained, no optimality proof under generic conditions has been provided
and no general discussion of efficiency and superefficiency in discrete
parameter models has appeared in the literature. In the present paper,
we provide a full answer to these problems in the case of discrete
parameter models for samples of i.i.d. (independent and identically distributed)
random variables. Therefore, after introducing some examples of discrete
parameter models in Section~\ref{Sect-Examples}, in Section~\ref{Sect-m-estimators}
we investigate the properties of a class of $m$-estimators. In particular,
in Section~\ref{Sect-ConsistencyofMLEandBE}, we derive some
conditions for strong consistency; then, in Section~\ref{Sect-AsymptoticDistributionoftheMLE},
we calculate an asymptotic approximation of the distribution of the
estimator and we establish its convergence rate. These results are
specialized to the case of the maximum likelihood estimator (\textsf{MLE})
and extended to Bayes estimators in Section~\ref{Sect-MLEandBayesEstimators}.
In Section~\ref{Sect-OptimalityandEfficiencyoftheMLE}, we
derive upper bounds for the convergence rate in the standard and in
the minimax contexts, and we discuss the relations between information
inequalities, efficiency and superefficiency. In particular, we prove
that estimators of discrete parameters have uncommon efficiency
properties. Indeed, under
the zero--one loss function, no estimator is efficient in the class
of consistent estimators for any value of $\theta_{0}\in\Theta$ ($\theta_{0}$
being here the true value of the parameter) and no estimator attains
the information inequality we derive. But the \textsf{MLE} still has
some appealing properties since it is minimax efficient and attains
the minimax information inequality bound.

\section{Examples of Discrete Parameter Models}\label{Sect-Examples}

The following examples are intended to show the relevance of discrete
parameter spaces in applied and theoretical statistics. In particular,
they show that the results in the following sections solve some long-standing
problems in statistics, optimization, information theory and signal
processing.

We recall that a \textit{statistical model} is a collection of probability
measures $\mathcal{P}=\{ \mathbb{P}_{\theta},\theta\in\Theta\} $
where $\Theta$ is the \textit{parameter space}. $\Theta$ is a subset
of a Euclidean or of a more abstract space.
\begin{example}[(Tumor transplantability)] We consider tumor transplantability
in mice. For a certain type of mating, the probability of a tumor
``taking'' when transplanted\vadjust{\goodbreak} from the grandparents to the offspring
is equal to $(\frac{3}{4})^{\theta}$ where $\theta$ is
an integer equal to the number of genes determining transplantability.
For another type of mating, the probability is $(\frac{1}{2})^{\theta}$.
We aim at estimating $\theta$ knowing that $n_{0}$ transplants take
out of $n.$ The likelihood is given by
\begin{eqnarray}
\ell_{n}(\theta)=\pmatrix{{n}\cr{n_{0}}}\cdot k^{\theta n_{0}}\cdot(1-k^{\theta
})^{n-n_{0}},\nonumber\\
\eqntext{\displaystyle\theta\in\mathbb{N},k\in\biggl\{ \frac{1}{2},\frac{3}{4}\biggr\}
.}
\end{eqnarray}
In this case the parameter space is discrete and the maximum likelihood
estimator can be shown to be $
\hat{\theta}^{n}=\operatorname{ni}[\frac{\ln({n_{0}}/{n})}{\ln k}]$
where $\operatorname{ni}[x]$ is the integer nearest to~$x$
(see \cite{Hammersley-JRSSB-1950}, page 236).
\end{example}
\begin{example}[(Exponential family restricted to a~lattice)] Consider a~random
variable $X$ distributed according to an exponential family where
the natural parameter $\theta$ is restricted to a lattice $\{ \theta
_{0}+\break\varepsilon\cdot N,N\in\mathbb{N}^{k}\} $,
for fixed $\theta_{0}$ and $\varepsilon$ (see \cite{Lindsay-Roeder-JASA-1987},
page 759). The case of a Gaussian distribution has been considered in
\cite{Hammersley-JRSSB-1950} (page~192) and \cite{Khan-AoS-1973,Khan-JSPI-2000},
the Poisson case in \cite{Hammersley-JRSSB-1950} (page 199), \cite
{McCabe-AoMS-1972,Stark-JASA-1975}.
In particular, \cite{Hammersley-JRSSB-1950} uses the Gaussian model
to estimate the molecular weight of insulin, assumed to be an integer
(however, see the remarks of Tweedie in the discussion of the same
paper).
\end{example}
\begin{example}[\hspace*{4pt}(Stochastic\hspace*{4pt} discrete\hspace*{4pt} optimization)] We consider the optimization
problem of the form $\min_{x\in S}g(x)$, where $g(x)=\mathbb{E}G(x,W)$
is an integral functional, $\mathbb{E}$ is the mean under probability
$\mathbb{P}$,\break $G(x,w)$ is a real-valued function of two
variables $x$ and $w$, $W$ is a random variable having probability
distribution $\mathbb{P}$ and $S$ is a finite set.\vspace*{1pt} We approximate
this problem through the sample average function $\hat
{g}_{n}(x)\triangleq\frac{1}{n}\sum_{i=1}^{n}G(x,W_{i})$
and the associated problem $\min_{x\in S}\hat{g}_{n}(x)$.
See \cite{Kleywegt-Shapiro-Homem-de-Mello-SIAMJO-2001} for some theoretical
results and a discussion of the stochastic knapsack problem and \cite
{vanderVlerk-WWW-2003}
for an up-to-date bibliography.
\end{example}
\begin{example}[(Approximate inference)] In many applied cases, the requirement
that the true model generating the data corresponds to a point belonging
to the parameter space appears to be too strong and unlikely. Moreover,
the objective is often to recover a model reproducing some stylized
facts from the original data. In these cases, approximation of a continuous
parameter space with a finite number of points allows for obtaining
such a model under weaker assumptions. This situation arises, for
example, in signal processing and automatic control applications\vadjust{\goodbreak} \cite
{Hawkes-Moore-IEEETAC-1976,Hawkes-Moore-IEEETIT-1976,Hawkes-Moore-PIEEE-1976,Baram-Sandell-BOOK-1977,Baram-Sandell-IEEETAC-1978a,Baram-Sandell-IEEETAC-1978b}
and is reminiscent of some related statistical techniques, such as
the \textit{discretization} device of Le Cam (\cite{LeCam-Yang-BOOK-1990},
Section 6.3), or the \textit{sieve estimation} of Grenander (\cite
{Grenander-BOOK-1981};
see also \cite{Geman-Hwang-AoS-1982}, Remark~5).
\end{example}
\begin{example}[($M$-ary hypotheses testing and related fields)] In information
theory, discrete parameter models are quite common, and their estimation
is a generalization of binary hypothesis testing that goes under the
names of $M$-\textit{ary} \textit{hypotheses} (or \textit{multihypothesis})
\textit{testing}, \textit{classification} or \textit{detection} (see the
examples in \cite{Nafie-Tewfik-BOOK-1998}). Consider a received waveform
$r(t)$ described by the equation $r(t)=m(t)+\sigma n(t)$
for $t\geq0$, where $m(t)$ is a deterministic signal,
$n(t)$ is an additive Gaussian white noise and~$\sigma$
is the noise intensity. The set of possible signals is restricted
to a finite number of alternatives, say $\{ m_{0}(t),\ldots,m_{J}(t)\} $:
the chosen signal is usually the one that maximizes the log-likelihood
of the sample, or an alternative criterion function. For example,
if the log-likelihood of the process based on the observation window
$[0,T]$ is used, we have
\begin{eqnarray*}
\hat{m}_{j}(\cdot)&=&\arg\max_{j=0,\ldots,J}\frac{1}{\sigma^{2}}\biggl[\int
_{0}^{T}m_{j}(t)r(t)\,\mathrm{d}t\\
&&\phantom{\arg\max_{j=0,\ldots,J}\frac{1}{\sigma^{2}}\biggl[}{}-\frac{1}{2}\int
_{0}^{T}m_{j}^{2}(t)\,\mathrm{d}t\biggr].
\end{eqnarray*}
Much more complex cases can be dealt with; see \cite{Hero-HoSP-1999}
for an introduction.
\end{example}

\section{$m$-Estimators in Discrete Parameter Models}\label{Sect-m-estimators}

In this section, we consider an estimator obtained by maximizing an objective
function of the form%
%
\[
Q_{n}(\theta)=\frac{1}{n}\sum_{i=1}^{n}\ln q(y_{i};\theta);
\]
in what follows, we allow for misspecification. Note that the
expression $m$-estimator stands for \textit{maximum likelihood type
estimator}, in the spirit of Huber~\cite{Huber-AoMS-1972},
and not for \textit{maximum} (or \textit{extremum}) \textit{estimator}
(see, e.g., \cite{Newey-McFadden-HoE-1994}, page 2114).

\subsection{Consistency of $m$-Estimators}\label{Sect-ConsistencyofMLEandBE}

In the case of a discrete parameter space, uniform convergence reduces
to pointwise convergence. Therefore, $m$-estimators are strongly
consistent under less stringent conditions than in the standard case;
in particular, no condition is needed on the continuity or differentiability
of the objective\vadjust{\goodbreak} function. The following assumption is used in order
to prove consistency in the case of i.i.d. replications:
\begin{enumerate}[A1.]
\item[A1.] The data $(Y_{i})_{i=1}^{n}$
are realizations of i.i.d. $(\frak{Y},\mathcal{Y})$-valued
random variables having probability measure $\mathbb{P}_{0}$.

The estimator $\hat{\theta}^{n}$ is obtained by maximizing over the
set $\Theta=\{ \theta_{0},\theta_{1},\ldots,\theta_{J}\} $,
of finite cardinality, the objective function
\[
Q_{n}(\theta)\triangleq\frac{1}{n}\sum_{i=1}^{n}\ln q(y_{i};\theta).
\]
The function $q$ is $\mathcal{Y}$-measurable for each $\theta\in\Theta$
and satisfies the $L^{1}$-domination condition\break $\mathbb{E}_{0}| \ln q(Y;\theta)|<+\infty$ for every $\theta\in\Theta$,
where $\mathbb{E}_{0}$ denotes the expectation taken under the true
probability measure $\mathbb{P}_{0}$.

Moreover, $\theta_{0}$ is the point of $\Theta$ maximizing $\mathbb
{E}_{0}\ln q(Y;\theta)$
and $\theta_{0}$ is globally identified (see~\cite
{Newey-McFadden-HoE-1994}, Section 2.2).
\end{enumerate}
\begin{rem}
(i) The assumption of a finite parameter space seems restrictive
with respect to the more general assumption of $\Theta$ being countable
(see, e.g., \cite{Hammersley-JRSSB-1950}). However, A1 is
compatible with the convex hull of $\Theta$ being compact, as in
standard asymptotic theory. Indeed, the cases analyzed in \cite
{Hammersley-JRSSB-1950}
have convex likelihood functions and this is a well-known substitute
for compactness of $\Theta$ (see 
\cite{Newey-McFadden-HoE-1994}, page 2133; see~\cite
{Choirat-Hess-Seri-AoP-2003},
for consistency with neither convexity nor compactness). Moreover,
the restriction to finite parameter spaces seems to be necessary to
derive the asymptotic approximation to the distribution of
$m$-estimators.\vspace*{-9pt}
\begin{longlist}[(ii)]
\item[(ii)]The relative position of the points of $\Theta$
is unimportant and the choice of $\theta_{0}$ as the maximizer is
arbitrary and is made only for practical purposes. Note that $\theta_{0}$
has no link with $\mathbb{P}_{0}$ apart from being the pseudo-true
value of $\ln q$ with respect to $\mathbb{P}_{0}$ on the parameter
space $\Theta$ (see, e.g., \cite{Gourieroux-Monfort-BOOK-1995}, Volume
1, page 14).
\end{longlist}
\end{rem}
\begin{prop}\label{Pr-ConsistencyoftheMLE}
Under Assumption \textup{A1},
the $m$-estimator $\hat{\theta}^{n}$ is a $\mathbb{P}_{0}$-strongly
consistent estimator of~$\theta_{0}$ and is $\mathcal{Y}^{\otimes
n}$-measurable.
\end{prop}
\begin{rem}
A similar result of consistency for discrete parameter spaces has
been provided by \cite{Silvey-JRSSB-1961} (page~446), by \cite
{Caines-AoS-1975,Caines-BOOK-1988}
(pages 325--333), by \cite{Barron-AoP-1985}\break (pages~1293--1294) as an application
of the Shannon--McMillan--Breiman Theorem of information theory, by
\cite{Wong-AoS-1986} (Section 2.1) as a preliminary result of his work
on partial likelihood, and by \cite{Manski-BOOK-1988} (page 96, Section
7.1.6).
\end{rem}

\subsection{Distribution of the $m$-Estimator}\label{Sect-AsymptoticDistributionoftheMLE}

For a discrete parameter space, the finite sample distribution of
the $m$-estimator $\hat{\theta}^{n}$ is a discrete distribution
converging to a Dirac mass concentrated at $\theta_{0}$. Since
the determination of an asymptotic approximation to this distribution
is an interesting and open problem, we derive in this section upper
and lower bounds and asymptotic estimates for probabilities of the
form $\mathbb{P}_{0}(\hat{\theta}^{n}=\theta_{i})$.

To simplify the following discussion, we introduce the processes:
%
%
\begin{eqnarray}\label{EqLikelihoodRatioProcesses}
\quad\cases{
Q_{n}(\theta_{j})\triangleq\displaystyle\frac{1}{n}\cdot\displaystyle\sum_{i=1}^{n}\ln
q(y_{i};\theta_{j}),\vspace*{2pt}\cr
\mathbf{X}_{k}^{(i)}\triangleq[\ln q(Y_{k};\theta_{i})\vspace*{2pt}\cr
\phantom{\mathbf{X}_{k}^{(i)}\triangleq[}{}-\ln
q(Y_{k};\theta_{j})]_{j=0,\ldots,J,j\neq i},\vspace*{2pt}\cr
\mathbf{X}_{k}\triangleq\mathbf{X}_{k}^{(0)}\vspace*{2pt}\cr
\phantom{\mathbf{X}_{k}}=[\ln q(Y_{k};\theta
_{0})-\ln q(Y_{k};\theta_{j})]_{j=1,\ldots,J},
}\\
\eqntext{i=1,\ldots,J,}
\end{eqnarray}

The probability of the estimator $\hat{\theta}^{n}$ taking on the
value $\theta_{i}$ can be written as
%
%
\begin{eqnarray}\label{Eq-ProbabilityforthetaiinLDForm}
\mathbb{P}_{0}(\hat{\theta}^{n}=\theta_{i})&=&\mathbb{P}_{0}\bigl(Q_{n}(\theta
_{i})>Q_{n}(\theta_{j}),\forall j\neq i\bigr)
\nonumber
\\[-8pt]
\\[-8pt]
\nonumber
&=&\mathbb{P}_{0}\Biggl(\sum
_{k=1}^{n}\mathbf{X}_{k}^{(i)}\in\operatorname{int }\mathbb{R}_{+}^{J}\Biggr).
\end{eqnarray}
The only approaches that have been successful in~our experience are
large deviations (in logarithmic and exact form) and saddlepoint approximations.
Note that we could have defined the probability
in~(\ref{Eq-ProbabilityforthetaiinLDForm})~as
$\mathbb{P}_{0}(Q_{n}(\theta_{i})\geq Q_{n}(\theta_{j}),\forall j\neq i)$
or through any other combination of equality and inequality signs;
this introduces some arbitrariness in the distribution of $\hat{\theta}^{n}$.
However, we will give some conditions (see Proposition~\ref{Pr-convergencerateofMLE2})
under which this difference is asymptotically irrelevant.

Section~\ref{Sect-APreliminaryResult} introduces definitions
and assumptions and discusses a preliminary result.
In Sec-\break tion~\ref{Sect-Largedeviationsasymptotics} we derive some results on the asymptotic behavior of $\mathbb
{P}_{0}(\hat{\theta}^{n}=\theta_{i})$
using large deviations principles (LDP). Then, we provide some
refinements of the previous expressions
using the theory of exact asymptotics for large deviations, with
special reference to the case $J=1$. At last, Section~\ref{Sect-SaddlepointApproximation} derives saddlepoint
approximations for probabilities of the form (\ref
{Eq-ProbabilityforthetaiinLDForm}).

\subsubsection{Definitions, assumptions and preliminary results}\label
{Sect-APreliminaryResult}

As concerns the distribution of the $m$-estima\-tor~$\hat{\theta}^{n}$,\vadjust{\goodbreak}
we shall need some concepts and functions derived from large deviations
theory (see \cite{Dembo-Zeitouni-BOOK-1998}); we recall that the
processes $Q_{n}(\theta_{j})$, $\mathbf{X}_{k}$ and $\mathbf{X}_{k}^{(i)}$
have been introduced in (\ref{EqLikelihoodRatioProcesses}).
Then, for $i=0,\dots,J$, we define the moment generating functions
\begin{eqnarray*}
M^{(i)}(\bolds{\lambda})&\triangleq&\mathbb{E}_{0}\bigl[e^{\sum
_{j=0,\ldots,J,j\neq i}\lambda_{j}\cdot[\ln q(Y;\theta_{i})-\ln q(Y;\theta
_{j})]}\bigr]\\
&=&\mathbb{E}_{0}\bigl[e^{\bolds{\lambda}^{\mathsf{T}}\mathbf{X}^{(i)}}\bigr],
\end{eqnarray*}
the logarithmic moment generating functions
\begin{eqnarray*}
\Lambda^{(i)}(\bolds{\lambda})&\triangleq&\ln M^{(i)}(\bolds{\lambda
})\\
&=&\ln\mathbb{E}_{0}\bigl[e^{\sum_{j=0,\ldots,J,j\neq i}\lambda_{j}\cdot[\ln
q(Y;\theta_{i})-\ln q(Y;\theta_{j})]}\bigr]\\
&=&\ln\mathbb{E}_{0}\bigl[e^{\bolds{\lambda}^{\mathsf{T}}\mathbf{X}^{(i)}}\bigr],
\end{eqnarray*}
and the Cram\'er transforms
\[
\Lambda^{(i),\ast}(\mathbf{y})\triangleq\sup_{\bolds{\lambda}\in\mathbb
{R}^{J}}\bigl[\langle\mathbf{y},\bolds{\lambda}\rangle-\Lambda
^{(i)}(\bolds{\lambda})\bigr],
\]
where $\langle\cdot,\cdot\rangle$ is the scalar product.
Note that, in what follows, $M(\bolds{\lambda})$, $\Lambda(\bolds
{\lambda})$
and $\Lambda^{\ast}(\mathbf{y})$ are respectively shortcuts
for $M^{(0)}(\bolds{\lambda})$, $\Lambda^{(0)}(\bolds{\lambda})$
and $\Lambda^{(0),\ast}(\mathbf{y})$. Moreover,
for a function $f\dvtx E\rightarrow\overline{\mathbb{R}}$, we will need
the definition of the \textit{effective domain} of $f$, $\mathcal
{D}_{f}\triangleq\{ x\in E\dvtx f(x)<\infty\} $.

The following assumptions will be used to approximate the distribution
of $\hat{\theta}^{n}$.
\begin{enumerate}[A2.]
\item[A2.] There exists a $\delta>0$ such that, for
any $\eta\in(-\delta,\delta)$, we have
\[
\mathbb{E}_{0}\biggl[\frac{q(Y;\theta_{j})}{q(Y;\theta_{k})}\biggr]^{\eta}<+\infty\quad
\forall j,k=0,\ldots,J.
\]
\end{enumerate}
%
%
\begin{rem}
$\!\!\!$In what follows, this assumption could be replaced by a condition
as in \cite{Ney-Robinson-JCA-1995} (Assumptions~H1 and H2).
\end{rem}
\begin{enumerate}[A3.]
\item[A3.] $\Lambda^{(i)}(\bolds{\lambda})$
is \textit{steep}, that is, $\lim_{n\rightarrow\infty}\Vert\frac
{\partial\Lambda^{(i)}(\mathbf{x})}{\partial\mathbf{x}}\Vert=\infty$
whenever $\{ \mathbf{x}_{n}\} _{n}$ is a sequence in $\operatorname{int
}(\mathcal{D}_{\Lambda^{(i)}})$
converging to a boundary point of $\operatorname{int }\mathcal{D}_{\Lambda
^{(i)}}$.
\end{enumerate}
\begin{rem} Under Assumptions \textup{A1}, \textup{A2}
and \textup{A3}, $\Lambda^{(i)}(\cdot)$
is \textit{essentially smooth} (see, e.g., \cite{Dembo-Zeitouni-BOOK-1998},
page 44). A sufficient condition for \textup{A3}
and essential smoothness is openness of $\mathcal{D}_{\Lambda^{(i)}}$
(see \cite{Ney-AoP-1984}, page 905, and~\cite{Iltis-JTP-1995}, pages
505--506).
\end{rem}

%
\begin{enumerate}[A4.]
\item[A4.] $\operatorname{int }(\mathbb{R}_{+}^{J}\cap\mathcal
{S}^{(i)})\neq\varnothing$,
where $\mathcal{S}^{(i)}$ is the closure of the convex
hull of the support of the law of $\mathbf{X}^{(i)}$.
\end{enumerate}
We will also need the following lemma showing the equivalence between Assumption
\textup{A2} and the so-called \textit{Cram\'er condition}
$\mathbf{0}\in\operatorname{int }(\mathcal{D}_{\Lambda^{(i)}})$,
for any $i=0,\ldots,J$.

\begin{lem}
\label{Lm-Cramerconditionandeta-Int}Under Assumption \textup{A1},
the following conditions are equivalent:
\begin{longlist}[(ii)]
\item[(i)] Assumption \textup{A2} holds;
\item[(ii)] $\mathbf{0}\in\operatorname{int }(\mathcal{D}_{\Lambda^{(i)}})$,
for any $i=0,\ldots,J$.\
\end{longlist}
\end{lem}

As concerns the saddlepoint approximation of Section~\ref{Sect-SaddlepointApproximation},
we need the following assumption:
\begin{enumerate}[A5.]
\item[A5.] The inequality
\begin{eqnarray*}
&&\biggl|\mathbb{E}_{0}\biggl[\prod_{j=0,\ldots,J,j\neq i}\biggl(\frac{q(Y;\theta
_{i})}{q(Y;\theta_{j})}\biggr)^{u_{j}+\iota\cdot t_{j}}\biggr]\biggr|\\
&&\quad<(1-\delta)\cdot
\biggl|\mathbb{E}_{0}\biggl[\prod_{j=0,\ldots,J,j\neq i}\biggl(\frac{q(Y;\theta
_{i})}{q(Y;\theta_{j})}\biggr)^{u_{j}}\biggr]\biggr|\\
&&\quad<\infty
\end{eqnarray*}
holds for $\mathbf{u}\in\operatorname{int }(\mathcal{D}_{\Lambda^{(i)}})$,
$\delta>0$ and $c<|\mathbf{t}|<\break C\cdot n^{{(s-3)}/{2}}$
($\iota$ denotes the imaginary unit). 
\end{enumerate}

\subsubsection{Large deviations asymptotics}\label
{Sect-Largedeviationsasymptotics}

In this section we consider large deviations asymptotics. We\break note
that, in what follows, $\operatorname{int }(\mathbb{R}_{+}^{J})^{c}$
stands for\break $\operatorname{int }\{ [(\mathbb{R}_{+})^{J}]^{c}\} $.
\begin{prop}\label{Pr-convergencerateofMLE2}\textup{(i)} For $i=1,\ldots,J$,
under Assumption \textup{A1}, the following result holds:
\[
\mathbb{P}_{0}(\hat{\theta}^{n}=\theta_{i})\geq\exp\Bigl\{ -n\cdot\inf
_{\mathbf{y}\in\operatorname{int }(\mathbb{R}_{+}^{J})}\Lambda^{(i),\ast
}(\mathbf{y})+o_{\inf}(n)\Bigr\},
\]
where $o_{\inf}(n)$ is a function such that\break $\liminf_{n\rightarrow\infty
}\frac{o_{\inf}(n)}{n}=0$.
\begin{longlist}[(iii)]
\item[(ii)] Under Assumptions \textup{A1} and \textup{A2}:
\[
\mathbb{P}_{0}(\hat{\theta}^{n}=\theta_{i})\leq\exp\Bigl\{ -n\cdot\inf
_{\mathbf{y}\in\mathbb{R}_{+}^{J}}\Lambda^{(i),\ast}(\mathbf{y})-o_{\sup
}(n)\Bigr\} ,
\]
where $o_{\sup}(n)$ is a function such that\break $\limsup_{n\rightarrow\infty
}\frac{o_{\sup}(n)}{n}=0$.\vspace*{6pt}

\item[(iii)] Under Assumptions \textup{A1}, \textup{A2}, \textup{A3} and \textup{A4}:
\begin{eqnarray*}
\mathbb{P}_{0}(\hat{\theta}^{n}=\theta_{i}) & =&\exp\Bigl\{ -\bigl(n+o(n)\bigr)\cdot\inf
_{\mathbf{y}\in\operatorname{int }(\mathbb{R}_{+}^{J})}\Lambda^{(i),\ast
}(\mathbf{y})\Bigr\} \\
& =&\exp\Bigl\{ -\bigl(n+o(n)\bigr)\cdot\inf_{\mathbf{y}\in\mathbb{R}_{+}^{J}}\Lambda
^{(i),\ast}(\mathbf{y})\Bigr\} .
\end{eqnarray*}
\end{longlist}
\end{prop}

\begin{prop}
\label{Pr-convergencerateofMLE}Under Assumption \textup{{A1}},
the following inequality holds:
\[
\mathbb{P}_{0}(\hat{\theta}^{n}\neq\theta_{0})\geq H\cdot\exp\Bigl\{ -n\cdot
\inf_{\mathbf{y}\in\operatorname{int }(\mathbb{R}_{+}^{J})^{c}}\Lambda^{\ast
}(\mathbf{y})+o_{\inf}(n)\Bigr\} ,
\]
where $H$ is the finite cardinality of the set\break $\arg\inf_{\mathbf{y}\in
\operatorname{int }(\mathbb{R}_{+}^{J})^{c}}\Lambda^{\ast}(\mathbf{y})$
and $o_{\inf}(n)$ is a function such that $\liminf_{n\rightarrow\infty
}\frac{o_{\inf}(n)}{n}=0$.

Under Assumptions \textup{A1} and \textup{A2}:
\[
\mathbb{P}_{0}(\hat{\theta}^{n}\neq\theta_{0})\leq H\cdot\exp\Bigl\{ -n\cdot
\inf_{\mathbf{y}\in\mathbb{R}_{+}^{J}}\Lambda^{\ast}(\mathbf{y})-o_{\sup
}(n)\Bigr\} ,
\]
where $o_{\sup}(n)$ is a function such that\break $\limsup_{n\rightarrow\infty
}\frac{o_{\sup}(n)}{n}=0$.
\end{prop}
\begin{rem}
The proposition allows us to obtain an upper bound on the bias of
the $m$-estimator, $\operatorname{\mathsf{Bias}}(\hat{\theta}^{n})\leq\sup_{j\neq
0}|\theta_{j}-\theta_{0}|\cdot\mathbb{P}_{0}(\hat{\theta}^{n}\neq\theta_{0})$.
\end{rem}

A better description of the asymptotic behavior of the probability
$\mathbb{P}_{0}(\hat{\theta}^{n}=\theta_{i})$
could be obtained, under some additional conditions, from the study
of the neighborhood of the contact point between the set $(\mathbb{R}_{+})^{J}$
and the level sets of the Cram\'er transform $\Lambda^{(i),\ast}(\cdot)$.
We leave the topic for future work. Here we just remark the following
brackets on the convergence rate.
\begin{prop}
\label{Pr-NeyconvergencerateofMLE} Under Assumptions \textup{A1},
\textup{A2}, \textup{A3}
and \textup{A4}, for sufficiently large
$n$, the following result holds:
\begin{eqnarray*}
c_{1}\frac{e^{-n\cdot\inf_{\mathbf{y}\in\mathbb{R}_{+}^{J}}\Lambda
^{(i),\ast}(\mathbf{y})}
}{n^{J/2}}&\leq&\mathbb{P}_{0}(\hat{\theta}^{n}=\theta_{i})\\
&\leq& c_{2}\frac
{e^{-n\cdot\inf_{\mathbf{y}\in\mathbb{R}_{+}^{J}}\Lambda^{(i),\ast
}(\mathbf{y})}
}{n^{1/2}}
\end{eqnarray*}
for $i=1,\ldots,J$ and for some $0<c_{1}\leq c_{2}<+\infty$.
\end{prop}

When $J=1$, a more precise convergence rate can be obtained under the following
assumption:
\begin{enumerate}[A6.]
\item[A6.] When $J\!=\!1$, there is a positive value $\mu\!\in\!\operatorname{int }(\mathcal{D}_{\Lambda^{(1)}})$
such that $\frac{\partial\Lambda^{(1)}(\lambda)}{\partial\lambda}|_{\lambda=\mu}=0$.
Moreover, the law of $\ln\frac{q(Y;\theta_{1})}{q(Y;\theta_{0})}$
is nonlattice (see \cite{Dembo-Zeitouni-BOOK-1998}, page~110).
\end{enumerate}
\begin{prop}
\label{Pr-ExactAsymptoticsinR}
$\!\!\!$Under Assumptions \textup{A1},
\textup{A2},~\textup{A3}, \textup{A4} and
\textup{A6},
with $\Theta=\{ \theta_{0},\theta_{1}\} $ and $J=1$,
we have
\begin{eqnarray*}
\mathbb{P}_{0}(\hat{\theta}^{n}=\theta_{1}) & =&\mathbb{P}_{0}(\hat
{\theta}^{n}\neq\theta_{0})\\
&=&\frac{e^{n\cdot\Lambda^{(1)}(\mu)}}{\mu\cdot
\sqrt{\Lambda^{(1),\prime\prime}(\mu)2\pi n}}\cdot\bigl(1+o(1)\bigr)\\
& =&\frac{e^{-n\cdot\Lambda^{(1),\ast}(0)}}{(\Lambda^{(1),\ast})^{\prime
}(0)}\cdot\sqrt{\frac{(\Lambda^{(1),\ast})^{\prime\prime}(0)}{2\pi
n}}\\
&&{}\cdot\bigl(1+o(1)\bigr).
\end{eqnarray*}
\end{prop}
\begin{rem}
A refinement of the previous asymptotic rates can be obtained using
results in \cite{Blackwell-Hodges-AoMS-1959,Bahadur-Rao-AoMS-1960}.
\end{rem}

\subsubsection{Saddlepoint approximation}\label{Sect-SaddlepointApproximation}

In this section we consider a different kind of approximation of the
probabilities $\mathbb{P}_{0}(\hat{\theta}^{n}=\theta_{i})$.\vadjust{\goodbreak}
\begin{thmm}
\label{Th-SaddlepointApproximation}
$\!\!\!\!$Under Assumptions \textup{
{A1}}, \textup{{A2}}
and~\textup{{A5}}, for $i\neq0$,
it is possible to choose $\mathbf{u}$ such that, for every $\mathbf
{v}\in[(\operatorname{int }\mathbb{R}_{+}^{J})\ominus\frac{\partial\Lambda
^{(i)}(\mathbf{u})}{\partial\mathbf{u}}]$,
$\mathbf{u}^{\mathsf{T}}\mathbf{v}\geq0$ and
\begin{eqnarray*}
\mathbb{P}_{0}(\hat{\theta}^{n}=\theta_{i}) & = & \exp\biggl(n\biggl[\Lambda
^{(i)}(\mathbf{u})-\mathbf{u}\cdot\frac{\partial\Lambda^{(i)}(\mathbf
{u})}{\partial\mathbf{u}}\biggr]\biggr)\\
& & {}\cdot\bigl[e_{s-3}\bigl(\mathbf{u},\operatorname{int }\mathbb{R}_{+}^{J}\ominus
\mathbb{E}_{0}\mathbf{X}^{(i)}\bigr)\\
&&\hspace*{9pt}\phantom{\bigl[}{}+\delta\bigl(\mathbf{u},\operatorname{int }\mathbb
{R}_{+}^{J}\ominus\mathbb{E}_{0}\mathbf{X}^{(i)}\bigr)\bigr],
\end{eqnarray*}
where
\begin{eqnarray*}
&& e_{s-3}\bigl(\mathbf{u},\operatorname{int }\mathbb{R}_{+}^{J}\ominus\mathbb
{E}_{0}\mathbf{X}^{(i)}\bigr)\\
 &&\quad =  \int_{\operatorname{int }\mathbb
{R}_{+}^{J}\ominus\frac{\partial\Lambda^{(i)}(\mathbf{u})}{\partial
\mathbf{u}}}\frac{\exp(-n\mathbf{u}\cdot\mathbf{y}-n\Vert\mathbf
{y}^{\ast}\Vert^{2}/2)}{(2\pi/n)^{J/2}\Delta^{1/2}}\\
& &\quad\phantom{=  \int_{\operatorname{int }\mathbb
{R}_{+}^{J}\ominus\frac{\partial\Lambda^{(i)}(\mathbf{u})}{\partial
\mathbf{u}}}}{} \cdot\Biggl[1+\sum_{i=1}^{s-3}n^{-i/2}Q_{i\mathbf{u}}\bigl(\sqrt{n}\mathbf
{y}^{\ast}\bigr)\Biggr]\,\mathrm{d}\mathbf{y},\\
&&Q_{\ell\mathbf{u}}(\mathbf{x}) \\
&&\quad =  \sum_{m=1}^{\ell}\frac{1}{m!}{\sum}^{\ast}{\sum}^{\ast\ast}\biggl(\frac{\kappa_{\nu_{1}n}\cdots\kappa_{\nu
_{m}n}}{\nu_{1}!\cdots\nu_{m}!}\biggr)\\
&&\qquad{}\cdot H_{I_{1}}(x_{1})\cdots
H_{I_{d}}(x_{d}),\\
&&\bigl|\delta\bigl(\mathbf{u},\operatorname{int }\mathbb{R}_{+}^{J}\ominus\mathbb
{E}_{0}\mathbf{X}^{(i)}\bigr)\bigr| \\
&&\quad\leq C\cdot n^{-{(s-2)}/{2}}
\end{eqnarray*}
and $\mathbf{V}=\frac{\partial^{2}\Lambda^{(i)}(\mathbf{u})}{\partial
\mathbf{u}^{2}}$,
$\mathbf{y}^{\ast}=\mathbf{V}^{-1/2}\mathbf{y}$, $\Vert\mathbf{y}^{\ast
}\Vert^{2}=\mathbf{y}^{\ast}\cdot\mathbf{y}^{\ast}=\mathbf{y}^{\mathsf
{T}}\mathbf{V}^{-1}\mathbf{y}$,
$\Delta=|\mathbf{V}|$, $H_{m}$ is the usual Hermite--Che\-byshev
polynomial of degree $m$, $\sum^{\ast}$ denotes the sum
over all $m$-tuples of positive integers $(j_{1},\ldots,j_{m})$
satisfying $j_{1}+\cdots+j_{m}=\ell$, $\sum^{\ast\ast}$
denotes the sum over all $m$-tuples $(\nu_{1},\ldots,\nu_{m})$
with $\nu_{i}=(\nu_{1i},\ldots,\nu_{di})$, satisfying
$(\nu_{1i}+\cdots+\nu_{di}=j_{i}+2,i=1,\ldots,m)$, and
$I_{h}=\nu_{h1}+\cdots+\nu_{hm},h=1,\ldots,d$. Note that $Q_{\ell\mathbf{u}}$
depends on $\mathbf{u}$ through the cumulants calculated at~$\mathbf{u}$.
\end{thmm}
\begin{rem}
The main question that this theorem leaves open is the
choice of the point $\mathbf{u}$. Usually this point is chosen as
a solution $\hat{\mathbf{u}}$ of $\mathbf{m}(\hat{\mathbf{u}})=\hat{\mathbf{x}}$;
this corresponds to a saddlepoint in $\kappa(\mathbf{u})$.
\cite{Daniels-AoMS-1954} (Section 6) and \cite{Lugannani-Rice-AAP-1980}
(page 480) give some conditions for $J=1$; \cite{Jensen-BOOK-1995}
(page 23) and \cite{Barndorff-Nielsen-BOOK-1978} (page 153) give conditions
for general $J$. \cite{Jing-Robinson-AoS-1994} suggests that the
most common solution is to choose $\hat{\mathbf{x}}$ and $\hat{\mathbf{u}}$
($\hat{\mathbf{x}}$ belonging to the boundary of $[\operatorname{int }\mathbb
{R}_{+}^{J}\ominus\mathbb{E}_{0}\mathbf{X}^{(i)}]$
and $\hat{\mathbf{u}}$ solving $\mathbf{m}(\hat{\mathbf{u}})=\hat{\mathbf{x}}$),
such that for every $\mathbf{v}\in[\operatorname{int }\mathbb
{R}_{+}^{J}\ominus\frac{\partial\Lambda^{(i)}(\mathbf{u})}{\partial
\mathbf{u}}]$,
$\hat{\mathbf{u}}^{\mathsf{T}}\mathbf{v}\geq0$. This is the same
as a dominating point in \cite{Ney-AoP-1983,Ney-AoP-1984,Ney-RES-1999};
therefore, \textup{{A2}}, \textup{{A3}}
and \textup{{A4}}, for sufficiently large
$n$, imply the existence of this point for any $i$.\looseness=1
\end{rem}

\subsection{The \textsf{MLE} and Bayes Estimators in Discrete Parameter
Models}\label{Sect-MLEandBayesEstimators}

In this section, we show how the previous results can be applied to
the \textsf{MLE} and Bayes estimators~under the zero--one loss function.
The \textsf{MLE} is defined~by
\begin{eqnarray*}
\hat{\theta}^{n}&\triangleq&\arg\max_{\theta\in\Theta}\prod
_{i=1}^{n}f_{Y_{i}}(y_{i};\theta_{k})\\
&=&\arg\max_{\theta\in\Theta}\biggl[\frac
{1}{n}\sum_{i=1}^{n}\ln f_{Y_{i}}(y_{i};\theta)\biggr].
\end{eqnarray*}
This corresponds to the \textit{minimum-error-probability estimate}
of \cite{Poor-Verdu-IEEETIT-1995} and to the \textit{Bayesian estimator}
of \cite{Vajda-AM-1971a,Vajda-AM-1971b}. On the other hand, using
the prior densities given by $\pi(\theta)$ for $\theta\in\Theta$,
the posterior densities of the Bayesian estimator are given by
\[
\mathbb{P}\{ \theta_{k}|\mathbf{Y}\} =\frac{\prod
_{i=1}^{n}f_{Y_{i}}(y_{i};\theta_{k})\pi(\theta_{k})}{\sum
_{j=0}^{J}\prod_{i=1}^{n}f_{Y_{i}}(y_{i};\theta_{j})\pi(\theta_{j})}.
\]
The Bayes estimator relative to zero--one loss $\check{\theta}^{n}$
(see Section~\ref{Sect-RiskFunctions} for a definition) is the
mode of the posterior distribution and is given by
\begin{eqnarray}\label{Eq-DefinitionoftheBayesEstimators}
\check{\theta}^{n} & \triangleq&\arg\max_{\theta\in\Theta}\ln\mathbb
{P}\{ \theta|\mathbf{Y}\}
\nonumber
\\[-8pt]
\\[-8pt]
\nonumber
& =&\arg\max_{\theta\in\Theta}\Biggl[\frac{1}{n}\sum_{i=1}^{n}\ln
f_{Y_{i}}(y_{i};\theta)+\frac{\ln\pi(\theta)}{n}\Biggr].\
\end{eqnarray}
Note that the \textsf{MLE} coincides with the Bayes estimator corresponding
to the uniform distribution $\pi(\theta)=(J+1)^{-1}$
for any $\theta\in\Theta$.

Assumption \textup{{A1}} can be replaced by the \mbox{following}
ones (where Assumptions \textup{{A8}} and \textup{{A9}}
entail that~the likelihood function is asymptotically maximized at~$\theta_{0}$
only):\looseness=-1
\begin{enumerate}[A7.]
\item[A7.] The parametric statistical model $\mathcal{P}$
is formed by a set of probability measures on a measurable space
$(\Omega,\mathcal{A})$
indexed by a parameter $\theta$ ranging over a parameter space $\Theta
=\{ \theta_{0},\theta_{1},\ldots,\theta_{J}\} $,
of finite cardinality. Let $(\frak{Y},\mathcal{Y})$ be a
measurable space and $\mu$ a positive $\sigma$-finite measure defined
on $(\frak{Y},\mathcal{Y})$ such that, for every $\theta\in\Theta$,
$\mathbb{P}_{\theta}$ is equivalent to $\mu$; the densities
$f_{Y}(Y;\theta)$
are $\mathcal{Y}$-measurable for each $\theta\in\Theta$.

The data $(Y_{i})_{i=1}^{n}$ are i.i.d. realizations from the
probability measure $\mathbb{P}_{0}$.
\item[A8.] The log density satisfies the
$L^{1}$-domination\break condition
$\mathbb{E}_{0}|\ln f_{Y}(Y;\theta_{i})|<+\infty$, for $\theta_{i}\in
\Theta$,\break
where~$\mathbb{E}_{0}$ denotes the expectation taken under the true
probability measure $\mathbb{P}_{0}$.
\item[A9.] $\theta_{0}$ is the point of $\Theta$ maximizing
$\mathbb{E}_{0}\ln f_{Y}(Y;\theta)$ and is globally identified.\vadjust{\goodbreak}
\end{enumerate}

In order to obtain the consistency of Bayes estimators, we need the
following assumption on the behavior of the prior distribution:
\begin{enumerate}[A10.]
\item[A10.] The prior distribution verifies $\pi(\theta)>0$
for any $\theta\in\Theta$.
\end{enumerate}
Proposition~\ref{Pr-ConsistencyoftheMLE} holds for the \textsf{MLE}
under Assumptions~\textup{{A7}}, \textup{{A8}}
and \textup{{A9}}, while for Bayes estimators~\textup{{A10}}
is required, too. Note that, under correct specification (i.e., when
the true parameter value belongs to~$\Theta$), a~standard Wald's
argument (see, e.g., Lem\-ma~2.2 in~\cite{Newey-McFadden-HoE-1994}, page
2124) shows that $\mathbb{E}_{\theta_{0}}\ln f_{Y}(Y;\theta)$
is maximized for $\theta=\theta_{0}$.

As concerns the distribution of the \textsf{MLE}, we have to consider
the case in which $q(y;\theta)$ is given by $f_{Y}(y; \theta)$,
$Q_{n}(\theta)$ by the log-likelihood function $\mathsf{L}_{n}(\theta)$,
and $\mathbf{X}_{k}$ and $\mathbf{X}_{k}^{(i)}$ by the
log-likelihood processes:
\[
\cases{
\mathsf{L}_{n}(\theta_{j})\triangleq\displaystyle\frac{1}{n}\cdot\sum_{i=1}^{n}\ln
f_{Y_{i}}(y_{i};\theta_{j}),\vspace*{2pt}\cr
\mathbf{X}_{k}^{(i)}\triangleq[\ln f_{Y_{k}}(Y_{k};\theta_{i})-\ln
f_{Y_{k}}(Y_{k};\theta_{j})]_{j=0,\ldots,J,j\neq i},\vspace*{2pt}\cr
\mathbf{X}_{k}\triangleq[\ln f_{Y_{k}}(Y_{k};\theta_{0})-\ln
f_{Y_{k}}(Y_{k};\theta_{j})]_{j=1,\ldots,J}.}
\]
Also $M(\bolds{\lambda})$ and $M^{(i)}(\bolds{\lambda})$
are consequently defined. Propositions~\ref{Pr-convergencerateofMLE2}
and~\ref{Pr-convergencerateofMLE} hold when Assumption \textup{A1}
is replaced by Assumptions~\textup{{A7}}, \textup{{A8}}
and \textup{{A9}}.

When the model is correctly specified, it is interesting to stress
an interpretation of the moment generating function in discrete parameter
models. We note that the moment generating functions can be written
as follows:
\begin{eqnarray}\label{Eq-MGFasanHellingerTransform}
M^{(i)}(\bolds{\lambda})& \triangleq&\mathbb{E}_{\theta_{0}}\bigl[e^{\sum
_{j=0,\ldots,J,j\neq i}\lambda_{j}\cdot[\ln f_{Y}(Y;\theta_{i})-\ln
f_{Y}(Y;\theta_{j})]}\bigr]\nonumber\\
& =&\int f_{Y}(y;\theta_{i})^{\sum_{j=0,\ldots,J,j\neq i}\lambda_{j}}
\nonumber
\\[-8pt]
\\[-8pt]
\nonumber
&&\phantom{\int}{}\cdot
\prod_{j=1,\ldots,J,j\neq i}f_{Y}(y;\theta_{j})^{-\lambda_{j}}\\
&&\hspace*{44pt}\qquad{}\cdot
f_{Y}(y;\theta_{0})^{1-\lambda_{0}}\mu(\mathrm{d}y).\nonumber
\end{eqnarray}
Therefore, in this case, the moment generating function $M^{(i)}(\bolds{\lambda})$
reduces to the so-called Hellinger transform $\mathsf{H}_{\bolds{\gamma
}}(\theta_{0},\ldots,\theta_{J})$
(see \cite{LeCam-Yang-BOOK-1990}, page 43) for a certain linear transformation
of $\bolds{\lambda}$ in $\bolds{\gamma}$:
\begin{eqnarray*}
&&\mathsf{H}_{\bolds{\gamma}}(\theta_{0},\ldots,\theta_{J})\\
&&\quad\triangleq\int
\prod_{j=0}^{J}[\mathbb{P}_{\theta_{j}}(\mathrm{d}y)]^{\gamma_{j}}\\
&&\quad=\int
\Biggl[\prod_{j=0}^{J}f_{Y}(y;\theta_{j})^{\gamma_{j}}\Biggr]\mu(\mathrm{d}y),\quad \sum
_{j=0}^{J}\gamma_{j}=1.
\end{eqnarray*}
Moreover, due to its convexity, $\mathsf{H}_{\bolds{\gamma}}(\theta
_{0},\ldots,\theta_{J})$
is surely finite for $\bolds{\gamma}$ belonging to the closed simplex
in $\mathbb{R}^{J+1}$.

Proposition~\ref{Pr-NeyconvergencerateofMLE} holds if Assumption
\textup{{A1}} is replaced by Assumptions \textup{{A7}},
\textup{{A8}} and \textup{{A9}}, and if \textup{{A2}}
and~\textup{{A3}} hold true. However,
Assumption \textup{{A4}} is unnecessary;
indeed, the fact that $\operatorname{int }(\mathbb{R}_{+}^{J}\cap\mathcal
{S}^{(i)})\neq\varnothing$
can be proved showing that $\mathbf{0}\in\operatorname{int} (\mathcal{S}^{(i)})$.
This is equivalent to the existence, for $j=1,\ldots,J,j\neq i$, of
two sets $A_{j}^{\ast}$ and~$A_{j}^{\ast\ast}$ of positive $\mu$-measure
and included in the support of~$Y$ such that, for $y_{j}^{\ast}\in
A_{j}^{\ast}$
and $y_{j}^{\ast\ast}\in A_{j}^{\ast\ast}$, $
f_{Y}(y_{j}^{\ast};\theta_{i})>f_{Y}(y_{j}^{\ast};\theta_{j})$ and
$f_{Y}(y_{j}^{\ast\ast};\theta_{i})<f_{Y}(y_{j}^{\ast\ast};\theta
_{j})$. This follows easily noting that these densities have to integrate
to~$1$, are almost surely (a.s.) different according to
Assumption \textup{{A9}}
and have the same support according to Assumption \textup{{A7}}.

In order to derive the distribution of Bayes estimators, we consider
Equation (\ref{Eq-DefinitionoftheBayesEstimators}) and we let
$\ln\bolds{\pi}^{(i)}\triangleq[\ln\frac{\pi(\theta_{i})}{\pi(\theta
_{j})}]_{j=0,\ldots,J,j\neq i}$.
Then, we can write
\begin{eqnarray*}
&&\mathbb{P}_{0}(\check{\theta}^{n}=\theta_{i}) \\
&&\quad =\mathbb{P}_{0}\Biggl(\sum
_{k=1}^{n}\mathbf{X}_{k}^{(i)}+\ln\bolds{\pi}^{(i)}\in\operatorname{int}
\mathbb{R}_{+}^{J}\Biggr)\\
&&\quad =\mathbb{P}_{0}\Biggl(\sum_{k=1}^{n}\mathbf{X}_{k}^{(i)}\in\prod
_{j=0,\ldots,J,j\neq i}\biggl(\ln\frac{\pi(\theta_{i})}{\pi(\theta_{j})},+\infty\biggr)\Biggr),
\end{eqnarray*}
and we can use the previous large deviations or saddlepoint formulas,
simply changing the set over which the $\inf$ is taken. However,
care is needed since both formulas hold under the assumption
\[
\mathbb{E}_{0}\mathbf{X}_{k}^{(i)}+\frac{1}{n}\cdot\ln\bolds{\pi
}^{(i)}\in\operatorname{int} (\mathbb{R}_{+}^{J})^{c}.
\]
In the case $J=1$, the similarity of these formulas with the corresponding
ones for a Neyman--Pearson test is striking; this revives the interpretation
of a Neyman--Pearson test as a Bayesian estimation problem. Therefore,
our analysis can be seen as a~(minor) extension of the theory of hypothesis
testing to a larger number of alternatives.

\section{Optimality and Efficiency}\label{Sect-OptimalityandEfficiencyoftheMLE}

In this section, we are interested in the problem of efficiency, with
special reference to maximum likelihood and Bayes estimators. In what
follows, we will suppose that the true parameter value belongs to
$\Theta$; this will be reflected in the probabilities that will be
written as $\mathbb{P}_{0}=\mathbb{P}_{\theta_{0}}$. Indeed, efficiency
statements for misspecified models are quite difficult to interpret.

In the statistics literature, efficiency (or superefficiency) can
be defined comparing the behavior of\vadjust{\goodbreak} the estimator with respect to
a lower bound or, alternatively, to a class of estimators. In the
continuous case, the two concepts almost coincide (despite superefficiency).
However, in the discrete case, the two concepts diverge dramatically
and we need more care in the derivation of the information inequalities
and in the statement of the efficiency properties.

An interesting problem concerns the choice of\break a~measure of efficiency
for the \textsf{MLE} in discrete parameter models: in his seminal
paper, Hammersley~\cite{Hammersley-JRSSB-1950} derives a
generalization of Cram\'er--Rao
inequality for the variance that is also valid when the parameter
space is countable. The same inequality has been derived, in slightly
more generality, in \cite{Chapman-Robbins-AoMS-1951,Blyth-Roberts-PSBSMSP-1972}.
However, this choice is well-suited only in cases in which the \textsf{MSE}
is a good measure of risk, for example, if the limiting distribution of the
normalized estimator is normal. Following the discussion by Lindley
in~\cite{Hammersley-JRSSB-1950}, we consider a different cost
function $\mathcal{C}_{1}(\theta,\theta_{0})$, whose risk
function is given by the probability of missclassification:
\begin{eqnarray*}
\mathcal{C}_{1}(\tilde{\theta}^{n},\theta_{0})&=&\mathsf{1}_{\{ \tilde
{\theta}^{n}\neq\theta_{0}\} }, \\
\mathcal{R}_{1}(\tilde{\theta
}^{n},\theta_{0})&=&\mathbb{P}_{\theta_{0}}(\tilde{\theta}^{n}\neq\theta_{0}).
\end{eqnarray*}
We also define the \textit{Bayes risk} (under the zero--one loss
function) associated with a prior distribution $\pi$ on the parameter
space $\Theta$. In particular, we consider the Bayes risk under the
risk function $\mathcal{R}_{1}(\tilde{\theta}^{n},\theta_{0})$
as
\[
r_{1}(\tilde{\theta}^{n},\pi)=\sum_{j=0}^{J}\pi(\theta_{j})\cdot\mathbb
{P}_{\theta_{j}}(\tilde{\theta}^{n}\neq\theta_{j}).
\]
If $\pi(\theta_{j})=(J+1)^{-1}$ we define $\mathbb{P}_{e}\triangleq
r_{1}(\tilde{\theta}^{n},\pi)$
as the \textit{average probability of error}. Note that this is indeed
the measure of error used by \cite{Vajda-AM-1971a,Vajda-AM-1971b}.

Using the risk function $\mathcal{R}_{1}$, in Section~\ref
{Sect-InformationInequalities} we derive
some information inequalities and we prove in Section~\ref{Sect-OptimalityandEfficiency} some optimality and efficiency
results for Bayes and \textsf{ML} estimators. In Section~\ref
{Sect-RiskFunctions} we briefly deal with alternative risk functions.

\subsection{Information Inequalities}\label{Sect-InformationInequalities}

This section contains lower bounds for the previously introduced risk
function $\mathcal{R}_{1}$. In the specific case of discrete
parameters, these generalize and unify the lower bounds proposed in
\cite
{Hammersley-JRSSB-1950,Chapman-Robbins-AoMS-1951,Kester-Kallenberg-AoS-1986,Hall-ISR-1989}.

In the following, first of all, a lower bound is proved and then a
minimax version of the same result is obtained. When needed, we will
refer to the
former as \textit{Chapman--Robbins lower bound} (and to the related efficiency
concept as \textit{Chapman--Robbins} \textit{efficiency}) since it recalls
the lower bound proposed\vadjust{\goodbreak} by these two authors in their 1951 paper, and
to the
latter as \textit{minimax Chapman--Robbins lower bound}.
Then, from these results, we derive a lower bound for the Bayes
risk.

\subsubsection{Lower bounds for the risk function $\mathcal
{R}_{1}$}\label{Sect-LowerBound}

The proposition of this section is intended to play the role of Cram\'er--Rao
and Chapman--Robbins lower bounds for the variance. It corresponds
essentially to Stein's Lemma in hypothesis testing. 
Moreover, a~version of the same bound for estimators respecting
(\ref{Eq-PartiallyConsistentEstimator}) is provided;
this corresponds to a similar result proposed in~\cite{Finesso-Liu-Narayan-IEEETIT-1996}
%
\begin{prop}\label{Pr-Bahadurlowerbound}Under Assumptions \textup{{A7}}
and \textup{{A9}}, for a strongly consistent estimator $\tilde
{\theta}^{n}$:
%
%
\begin{eqnarray}\label{Eq-Bahadurbound-KLdivergence}
&&\lim_{n\rightarrow\infty}\frac{1}{n}\ln\mathcal{R}_{1}(\tilde{\theta
}^{n},\theta_{0})
\nonumber
\\[-8pt]
\\[-8pt]
\nonumber
&&\quad \ge \sup_{\theta_{1}\in\Theta\setminus\{ \theta
_{0}\} }\mathbb{E}_{\mathbb{\theta}_{1}}\ln\biggl(\frac{f_{Y}(Y;\theta
_{0})}{f_{Y}(Y;\theta_{1})}\biggr).
\end{eqnarray}
On the other hand, if
%
%
\begin{equation}\label{Eq-PartiallyConsistentEstimator}
\limsup_{n\rightarrow\infty}\mathbb{P}_{\theta_{j}}\{ \tilde{\theta
}^{n}\neq\theta_{j}\} <1,
\end{equation}
then
\[
\liminf_{n\rightarrow\infty}\frac{1}{n}\ln\mathcal{R}_{1}(\tilde{\theta
}^{n},\theta_{0})\ge\sup_{\theta_{1}\in\Theta\setminus\{ \theta_{0}\}
}\mathbb{E}_{\mathbb{\theta}_{1}}\ln\biggl(\frac{f_{Y}(Y;\theta
_{0})}{f_{Y}(Y;\theta_{1})}\biggr).
\]
\end{prop}
\begin{rem}
\label{Rm-IRoptimalityandCRoptimality}\textup{(i)} Note that
this inequality only holds for estimators that are consistent or respect
condition~(\ref{Eq-PartiallyConsistentEstimator}), while the
one of Proposition~\ref{Pr-Chapman-Robbinslowerbound} holds for
any estimator.\vspace*{-6pt}
\begin{longlist}[(ii)]
\item[(ii)] Proposition~\ref{Pr-Bahadurlowerbound} provides an
upper bound for the \textit{inaccuracy rate} of \cite
{Kester-Kallenberg-AoS-1986}:
\[
e(\varepsilon,\theta_{0},\tilde{\theta}^{n})  \leq\inf_{\theta_{1}\in
\Theta\setminus\{ \theta_{0}\} }\mathbb{E}_{\mathbb{\theta}_{1}}\ln
\biggl(\frac{f_{Y}(Y;\theta_{1})}{f_{Y}(Y;\theta_{0})}\biggr)
\]
for any $\varepsilon$ small enough ($\varepsilon<\min_{\theta_{1}\in
\Theta\setminus\{ \theta_{0}\} }\Vert\theta_{1}-\theta_{0}\Vert$).
\end{longlist}
\end{rem}

\subsubsection{Minimax lower bounds for the risk function~$\mathcal{R}_{1}$}\label{Sect-MinimaxLowerBound}

The following result is a minimax lower bound on the probability of
misclassification. It is based on the Neyman--Pearson Lemma and Chernoff's
Bound.
\begin{prop}
\label{Pr-Chapman-Robbinslowerbound}Under Assumptions \textup{A7}
and \textup{{A9}}, for any estimator $\tilde{\theta}^{n}$:
%
%
\begin{eqnarray}\label{Eq-ChernoffsLowerBound}
& & \hspace*{4pt}\liminf_{n\rightarrow\infty}\frac{1}{n}\ln\sup_{\theta_{0}\in\Theta
}\mathcal{R}_{1}(\tilde{\theta}^{n},\theta_{0})\nonumber\\
&&\hspace*{6pt}\quad \ge \sup_{\theta_{1}\in\Theta\setminus\{ \theta_{0}\} }\sup
_{\theta_{0}\in\Theta}\ln\biggl[\inf_{1>u>0}\int f_{Y}(y;\theta_{1})^{u}\\
&&\hspace*{124pt}\quad{}\cdot
f_{Y}(y;\theta_{0})^{1-u}\mu(\mathrm{d}y)\biggr].\nonumber
\end{eqnarray}
\end{prop}
\begin{rem} \label{Rm-HallBound}
\textup{(i)} The previous proposition provides an expression
for the \textit{minimax Bahadur risk} (also called \textit{(minimax) rate
of inaccuracy}; see \mbox{\cite{Bahadur-San-1960,Korostelev-Leonov-PPI-1996}})
analogous to Chernoff's Bound, thus providing a~minimax version of
Remark~\ref{Rm-IRoptimalityandCRoptimality}\textup{(ii)}.\vspace*{-6pt}

\begin{longlist}[(iii)]
\item[(ii)] Other methods to derive similar minimax inequalities
are Fano's Inequality and Assouad's Lem\-ma (see \cite
{LeCam-Yang-BOOK-1990}, page 220); however, in the present case they do
not allow us
to obtain tight bounds
, since the usual application
of these methods relies on the approximation of the parameter space
with a finite set of points $\Theta$ whose cardinality increases
with~$n$. Clearly, this cannot be done in the present case.

\item[(iii)] Using Lemma 5.2 in \cite{Puhalskii-Spokoiny-Ber-1998},
it is possible to show that the minimax bound is larger than the
classical one.
\item[(iv)]
Under Assumption \textup{{A10}}, the Bayes risk $r_{1}$ under
the risk function $\mathcal{R}_{1}$
and the prior $\pi$ respects the equality
%
%
\begin{equation}
\lim_{n\rightarrow\infty}\frac{1}{n}\ln r_{1}(\tilde{\theta}^{n},\pi
)=\lim_{n\rightarrow\infty}\frac{1}{n}\ln\max_{\theta_{0}\in\Theta
}\mathcal{R}_{1}(\tilde{\theta}^{n},\theta_{0}).\hspace*{-25pt}
\label{Eq-AsymptoticEquivalenceofBayesRiskandMinimaxProbError}
\end{equation}
Then, Proposition~\ref{Pr-Chapman-Robbinslowerbound} holds also
for the Bayes risk: clearly this bound is independent of the prior
distribution $\pi$ (provided it is strictly positive, i.e., \textup{A10}
holds) and also holds for the probability of error $\mathbb{P}_{e}$.
This inequality can be seen as an asymptotic version of the
van Trees inequality 
for a different risk function.
\end{longlist}
\end{rem}

\subsection{Optimality and Efficiency}\label{Sect-OptimalityandEfficiency}

In this section, we establish some optimality results for the \textsf{MLE}
in discrete parameter models. The situation is much more intricate than
in regular statistical models
under the quadratic loss function, in which efficiency coincides with
the attainment of the Cram\'er--Rao lower bound (despite superefficiency).
Therefore, we propose the following definition. We denote by $\mathcal
{R}=\mathcal{R}(\bar{\theta}^{n},\theta_{0})$
the risk function of the estimator $\bar{\theta}^{n}$ evaluated at
$\theta_{0}$, and by $\tilde{\Theta}$ a class of estimators.
\begin{defn}
The estimator $\bar{\theta}^{n}$ is \textit{efficient with respect
to} (w.r.t.) $\tilde{\Theta}$ \textit{and} w.r.t. $\mathcal{R}$
\textit{at} $\theta_{0}$ if
%
%
\begin{equation}
\mathcal{R}(\bar{\theta}^{n},\theta_{0})\leq\mathcal{R}(\tilde{\theta
}^{n},\theta_{0})\quad \forall\tilde{\theta}^{n}\in\tilde{\Theta}.\label
{Eq-DefinitionofEfficiency}
\end{equation}
The estimator $\bar{\theta}^{n}$ is \textit{minimax efficient} w.r.t.
$\tilde{\Theta}$ \textit{and} w.r.t. $\mathcal{R}$ if
%
%
\begin{equation}
\sup_{\theta_{0}\in\Theta}\mathcal{R}(\bar{\theta}^{n},\theta_{0})\leq
\sup_{\theta_{0}\in\Theta}\mathcal{R}(\tilde{\theta}^{n},\theta_{0})\quad
\forall\tilde{\theta}^{n}\in\tilde{\Theta}.\label
{Eq-DefinitionofMinimaxEfficiency}
\end{equation}
The estimator $\bar{\theta}^{n}$ is \textit{superefficient} w.r.t.
$\tilde{\Theta}$ \textit{and}\break w.r.t.~$\mathcal{R}$ if for every
$\tilde{\theta}^{n}\in\tilde{\Theta}$:
\[
\mathcal{R}(\bar{\theta}^{n},\theta_{0})\le\mathcal{R}(\tilde{\theta
}^{n},\theta_{0})\vadjust{\goodbreak}
\]
for every $\theta_{0}\in\Theta$ and there exists at least a value
$\theta_{0}^{*}\in\Theta$ such that the inequality is replaced by
a strict inequality for $\theta_{0}=\theta_{0}^{*}$.

The estimator $\bar{\theta}^{n}$ is \textit{asymptotically} $\mathrm
{CR}$-\textit{efficient}
w.r.t. $\mathcal{R}$ \textit{at} $\theta_{0}$ if it attains
the Chapman--Robbins lower bound of Proposition~\ref{Pr-Bahadurlowerbound}
at $\theta_{0}$ [say $\mathrm{CR-}\mathcal{R}(\theta_{0})$]
in the asymptotic form:\vspace*{1pt}
\[
\liminf_{n\rightarrow\infty}\frac{1}{n}\ln\mathcal{R}(\bar{\theta
}^{n},\theta_{0})=\ln\mathrm{CR-}\mathcal{R}(\theta_{0}).\vspace*{1pt}
\]
The estimator $\bar{\theta}^{n}$ is \textit{asymptotically minimax}
$\mathrm{CR}$-\textit{efficient} w.r.t. $\mathcal{R}$ if it
attains the minimax Chapman--Robbins lower bound of Proposition~\ref{Pr-Chapman-Robbinslowerbound}
(say $\mathrm{CR}\mbox{-}\mathcal{R}_{\max}$) in the asymptotic form:\vspace*{1pt}
\[
\liminf_{n\rightarrow\infty}\frac{1}{n}\ln\sup_{\theta_{0}\in\Theta
}\mathcal{R}(\bar{\theta}^{n},\theta_{0})=\ln\mathrm{CR-}\mathcal
{R}_{\max}.\vspace*{1pt}
\]
The estimator $\bar{\theta}^{n}$ is \textit{asymptotically} $\mathrm
{CR}$-\textit{superefficient}
w.r.t. $\mathcal{R}$ if\vspace*{1pt}
\[
\liminf_{n\rightarrow\infty}\frac{1}{n}\ln\mathcal{R}(\bar{\theta
}^{n},\theta_{0})\le\ln\mathrm{CR-}\mathcal{R}(\theta_{0})\vspace*{1pt}
\]
for every $\theta_{0}\in\Theta$ and there exists at least a value
$\theta_{0}^{*}\in\Theta$ such that the inequality is replaced by
a strict inequality for $\theta_{0}=\theta_{0}^{*}$.
\end{defn}
\begin{rem} As in Remark~\ref{Rm-IRoptimalityandCRoptimality}\textup{(ii)}, it is easy to~see that $\mathrm
{IR}$-optimality and
$\mathrm{CR}$-efficiency w.r.t. $\mathcal{R}_{1}$ coincide.
\end{rem}

The efficiency landscape offered by discrete parameter models will
be illustrated by Example~\ref{Exam-IntegerMeanofaGaussianSample}.
This shows that, even in the simplest case, that is, the estimation of
the integer mean of a Gaussian random variable with known variance,
the \textsf{MLE} does not attain the lower bound on the missclassification
probability but it attains the minimax lower bound. Moreover, simple
estimators are built that outperform the \textsf{MLE} for certain
values of the true parameter value $\theta_{0}$.
\begin{example}
\label{Exam-IntegerMeanofaGaussianSample}Let us consider the
estimation of the mean of a Gaussian distribution whose variance~$\sigma^{2}$
is known: we suppose that the true mean is~$\alpha$, while the parameter
space is $\{ -\alpha,\alpha\} $, where~$\alpha$ is known. The maximum
likelihood estimator $\hat{\theta}^{n}$ takes the
value $-\alpha$ if the sample mean takes on its value in $(-\infty,0)$
and $\alpha$ if it falls in $[0,+\infty)$ (the position
of $0$ is a convention). Therefore:\vspace*{1pt}
\begin{eqnarray*}
\mathbb{P}_{\theta_{0}}(\hat{\theta}^{n}\neq\theta_{0}) & =&\mathbb
{P}_{\theta_{0}}(\hat{\theta}^{n}=-\alpha)\\[3pt]
&=&\int_{-\infty}^{0}\frac
{e^{-{(\bar{y}-\alpha)^{2}}/{(2\sigma^{2}/n)}}}{\sqrt{2\pi\sigma
^{2}/n}}\,\mathrm{d}\bar{y}\\
&=&\int_{-\infty}^{-{\sqrt{n}\alpha}/{\sigma
}}\frac{e^{-{t^{2}}/{2}}}{\sqrt{2\pi}}\,\mathrm{d}t\\
& =&\Phi\biggl(-\frac{\sqrt{n}\alpha}{\sigma}\biggr)\\
&=&\frac{e^{-{n\alpha
^{2}}/{(2\sigma^{2})}}}{\sqrt{2\pi n}}\frac{\sigma}{\alpha}\cdot\biggl(1+O\biggl(\frac{1}{n}\biggr)\biggr),
\end{eqnarray*}
where we have used Problem 1 on page 193 in \cite{Feller-Book-1968}.
Proposition~\ref{Pr-ExactAsymptoticsinR} allows also for recovering
the right convergence rate. Indeed, we have 
%
\begin{eqnarray*}
\mathbb{P}_{\theta_{0}}(\hat{\theta}^{n}\neq\alpha)&=&\mathbb{P}_{\theta
_{0}}(\hat{\theta}^{n}=-\alpha)\\
&=&\frac{e^{-{n\alpha^{2}}/{(2\sigma
^{2})}}}{\sqrt{2\pi n}}\frac{\sigma}{\alpha}\cdot\bigl(1+o(1)\bigr).
\end{eqnarray*}
On the other hand, the lower bound of Proposition~\ref{Pr-Bahadurlowerbound}
yields
\[
\lim_{n\rightarrow\infty}\frac{1}{n}\ln\mathbb{P}_{\theta_{0}}(\hat
{\theta}^{n}\neq\theta_{0})\geq-\frac{2 \alpha^{2}}{\sigma^{2}},
\]
and the lower bound of Proposition~\ref{Pr-Chapman-Robbinslowerbound}
yields
\[
\liminf_{n\rightarrow\infty}\frac{1}{n}\sup_{\theta_{0}\in\{ -\alpha
,\alpha\} }\ln\mathbb{P}_{\theta_{0}}(\hat{\theta}^{n}\neq\theta
_{0})\geq-\frac{\alpha^{2}}{2\sigma^{2}}.
\]
Therefore, the \textsf{MLE} asymptotically attains the minimax lower
bound but not
the classical one.

In the following, we will show that estimators can be pointwise
more efficient than the \textsf{MLE}; consider the estimator defined
by
\[
\tilde{\theta}^{n}(k)=\cases{
\theta_{0} &  $\mathrm{if}\ \mathsf{L}_{n}(\theta_{0})\geq\mathsf
{L}_{n}(\theta_{1})+k\cdot n,$\vspace*{2pt}\cr
\theta_{1} &  $\mathrm{else.}$}
\]
When $k=0$, $\tilde{\theta}^{n}(k)$ coincides with the \textsf{MLE}
$\hat{\theta}^{n}$. Then, the behavior of the estimator is
characterized by the probabilities:
\begin{eqnarray*}
\mathbb{P}_{\theta_{0}}\bigl(\tilde{\theta}^{n}(k)=\theta_{0}\bigr) & =&\Phi\biggl(\frac
{k\cdot n\cdot\sigma^{2}+2\alpha^{2}\cdot n}{2\alpha\sigma\sqrt{n}}\biggr),\\
\mathbb{P}_{\theta_{1}}\bigl(\tilde{\theta}^{n}(k)=\theta_{0}\bigr) & =&\Phi\biggl(\frac
{k\cdot n\cdot\sigma^{2}-2\alpha^{2}\cdot n}{2\alpha\sigma\sqrt{n}}\biggr).
\end{eqnarray*}
We have (weak) consistency if
%
%
\begin{equation}
2\biggl(\frac{\alpha}{\sigma}\biggr)^{2}>k>-2\biggl(\frac{\alpha}{\sigma}\biggr)^{2}.\label
{Eq-ConsistencyConditionsintheExample}
\end{equation}
The risk $\mathcal{R}_{1}(\tilde{\theta}^{n}(k),\theta_{0})$
under $\theta_{0}$ is then
\[
\mathbb{P}_{\theta_{0}}\bigl(\tilde{\theta}^{n}(k)\neq\theta_{0}\bigr)  =  \Phi
\biggl[-\frac{k\cdot\sigma^{2}+2\alpha^{2}}{2\alpha\sigma}\cdot\sqrt{n}\biggr];
\]
this can be made smaller than the probability of error of the \textsf
{MLE} simply taking $k>0$,
thus implying that the \textsf{MLE} is not pointwise efficient.\vadjust{\goodbreak}

Now, we show that this estimator cannot converge faster than the
Chapman--Robbins lower bound without losing its consistency. Indeed,
$\mathbb{P}_{\theta_{0}}(\tilde{\theta}^{n}(k)\neq\theta_{0})$
is smaller than the Chapman--Robbins lower bound if
\[
k^{2}+4k\biggl(\frac{\alpha}{\sigma}\biggr)^{2}-12\biggl(\frac{\alpha}{\sigma}\biggr)^{4}\geq0,
\]
and this is never true under (\ref{Eq-ConsistencyConditionsintheExample}).
If this estimator is pointwise more efficient than the \textsf{MLE}
under $\theta_{0}$, then its risk under $\theta_{1}$ is given by
\[
\mathbb{P}_{\theta_{1}}\bigl(\tilde{\theta}^{n}(k)\neq\theta_{1}\bigr)=\Phi\biggl[\frac
{k\cdot\sigma^{2}-2\alpha^{2}}{2\alpha\sigma}\cdot\sqrt{n}\biggr],
\]
and this is greater than for the \textsf{MLE}. This shows that a
faster convergence rate can be obtained in some points, the price
to pay being a worse convergence rate elsewhere in $\Theta$.
\end{example}

\subsubsection{Optimality w.r.t. classes of estimators}\label{Sect-OptimalityofBayesandMLEs}

In the following section, we show some optimality properties of Bayes
and \textsf{ML} estimators. We start with an important and well-known
fact.
\begin{prop}
\label{Pr-MinimizationoftheBayesRisk}Under \textup{{A7}},
\textup{{A8}}, \textup{{A9}} and \textup{{A10}},
the Bayes risk $r_{1}(\tilde{\theta}^{n},\pi)$ (under
the zero--one loss function) associated with a prior distribution $\pi$
is strictly minimized by the posterior mode corresponding to the prior
$\pi$, for any finite $n$.
\end{prop}

The following proposition shows that the \textsf{MLE} is admissible
and minimax efficient under the zero--one loss and minimizes the average
probability
of error. It implies that estimators
that are more efficient than the \textsf{MLE} at a certain point $\theta
_{0}\in\Theta$
are less efficient in at least another point $\theta_{1}\in\Theta$.
As a~result, estimators can
be more efficient than minimax efficient ones only on portions of
the parameter space, but are then strictly less efficient elsewhere.
\begin{prop}
\label{Pr-AdmissibilityandMinimaxEfficiencyofMLE}Under Assumptions
\textup{{A7}}, \textup{{A8}}\break and~\textup{{A9}},
the \textsf{MLE} is admissible and minimax efficient w.r.t.
the class of all estimators and w.r.t. $\mathcal{R}_{1}$ and
minimizes the average probability of error $\mathbb{P}_{e}$.
\end{prop}

\subsubsection{Optimality w.r.t. the information inequalities}\label
{Sect-OptimalitywrtInformationInequalities}

In this subsection, we will show that the \textsf{MLE} does not attain
the Chapman--Robbins lower bound in the form of Proposition~\ref{Pr-Bahadurlowerbound}
but that it attains the minimax form of Proposition~\ref{Pr-Chapman-Robbinslowerbound}
and that efficiency and minimax efficiency are generally incompatible.

Therefore, the situation described in Example~\ref{Exam-IntegerMeanofaGaussianSample}
is general, for it is possible to show that the \textsf{MLE} is generally
inefficient with respect to the lower bounds exposed in Proposition
\ref{Pr-Bahadurlowerbound}.
\begin{prop}
\label{Pr-Proofofnon-IR-optimality}Under Assumptions \textup{{A7}},
\textup{{A8}} and \textup{{A9}}:
\begin{longlist}[(iii)]
\item[(i)] the \textsf{MLE} is not asymptotically $\mathrm{CR}$-\break efficient
w.r.t. $\mathcal{R}_{1}$ at $\theta_{0}$;

\item[(ii)] the \textsf{MLE} is asymptotically minimax $\mathrm{CR}$-\break efficient
w.r.t. $\mathcal{R}_{1}$;

\item[(iii)] an estimator that is asymptotically $\mathrm{CR}$-\break efficient
w.r.t. $\mathcal{R}_{1}$ at $\theta_{0}$ is not asymptotically
minimax $\mathrm{CR}$-efficient w.r.t. $\mathcal{R}_{1}$.
\end{longlist}
\end{prop}
\begin{rem}
The assumption of homogeneity of the probability measures, necessary to
derive \textup{(ii)},
can be removed in the proof of \textup{(i)} along the lines of~\cite
{Kester-Kallenberg-AoS-1986}.
\end{rem}

\subsubsection{The evil of superefficiency}\label{Sect-Superefficiency}

Ever since it was discovered by Hodges, the problem of superefficiency
has been dealt with extensively in regular statistical problems (see,
e.g., \cite{LeCam-UCPS-1953,vanderVaart-BOOK-1997}). However, these
proofs do not transpose to discrete parameter estimation problems,
since they are mostly based on the equivalence of prior probability
measures with the Lebesgue measure and on properties of Bayes estimators
that do not hold in this case. Moreover, the discussion of the previous
sections has shown that, in discrete parameter problems, $\mathrm
{CR}$-efficiency
and efficiency with respect to a class of estimators do not coincide.
The following proposition yields a~solution to the superefficiency
problem.
\begin{prop}
\label{Pr-Superefficiency}Under Assumptions \textup{{A7}},
\textup{{A8}} and \textup{{A9}}:
\begin{longlist}
\item[(i)] no estimator $\tilde{\theta}^{n}$ is asymptotically
$\mathrm{CR}$-super\-efficient w.r.t. $\mathcal{R}_{1}$ at $\theta
_{0}\in\Theta$;

\item[(ii)] no estimator $\tilde{\theta}^{n}$ is superefficient
w.r.t. the \textsf{MLE} and $\mathcal{R}_{1}$.
\end{longlist}
\end{prop}

\subsection{Alternative Risk Functions}\label{Sect-RiskFunctions}

Now we consider in what measure the previous results transpose when
changing the risk function.
Following \cite{Hammersley-JRSSB-1950}, we first consider the quadratic
cost function and the corresponding risk function:
\begin{eqnarray*}
\mathcal{C}_{2}(\tilde{\theta}^{n},\theta_{0})&=&(\tilde{\theta
}^{n}-\theta_{0})^{2}, \\
 \mathcal{R}_{2}(\tilde{\theta}^{n},\theta
_{0})&=&\operatorname{\mathsf{MSE}}(\tilde{\theta}^{n}).
\end{eqnarray*}
The cost function $\mathcal{C}_{1}$ has the drawback of weighting in
the same
way points of the parameter space that lie at different distances
with respect to the true value $\theta_{0}$. In many cases, a more
general loss function can be considered, as suggested in \cite
{Gourieroux-Monfort-BOOK-1995}
(Volume 1, page 51) for multiple tests:
\[
\mathcal{C}_{3}(\tilde{\theta}^{n},\theta_{0})=\cases{
0 & $\mathrm{if}\ \tilde{\theta}^{n}=\theta_{0},$\vspace*{1pt}\cr
a_{j}(\theta_{0}) & $\mathrm{if}\ \tilde{\theta}^{n}=\theta_{j},$}\vadjust{\goodbreak}
\]
where $a_{j}(\theta_{0})>0$ for $j=1,\ldots,J$ can be tuned
in order to give more or less weight to different points of the parameter
space. The risk function is therefore given by the weighted probability
of misclassification $\mathcal{R}_{3}(\tilde{\theta}^{n},\theta
_{0})=\sum_{j=1}^{J}a_{j}(\theta_{0})\cdot\mathbb{P}_{\theta_{0}}\{
\tilde{\theta}^{n}=\theta_{j}\} $.

It is trivial to remark that
\begin{eqnarray*}
&&\lim_{n\rightarrow\infty}\frac{1}{n}\ln\mathcal{R}_{2}(\tilde{\theta
}^{n},\theta_{0}) \\
&&\quad = \lim_{n\rightarrow\infty}\frac{1}{n}\ln\mathbb
{P}_{\theta_{0}}(\tilde{\theta}^{n}\neq\theta_{0}),\\
&&\liminf_{n\rightarrow\infty}\frac{1}{n}\ln\sup_{\theta_{0}\in\Theta
}\mathcal{R}_{2}(\tilde{\theta}^{n},\theta_{0}) \\
&&\quad =  \liminf
_{n\rightarrow\infty}\frac{1}{n}\ln\sup_{\theta_{0}\in\Theta}\mathbb
{P}_{\theta_{0}}(\tilde{\theta}^{n}\neq\theta_{0}),
\end{eqnarray*}
and the lower bounds of Propositions~\ref{Pr-Bahadurlowerbound}
and~\ref{Pr-Chapman-Robbinslowerbound} hold also in this case.
The same equalities hold also for~$\mathcal{R}_{3}$. As a result,
Proposition~\ref{Pr-Proofofnon-IR-optimality} and Proposition~\ref{Pr-Superefficiency}\textup{(i)} apply also to these risk functions.

On the other hand, as concerns Proposition~\ref{Pr-AdmissibilityandMinimaxEfficiencyofMLE} and Proposition~\ref{Pr-Superefficiency}\textup{(ii)}, it is simple to show that with
respect to
the risk functions $\mathcal{R}_{2}(\tilde{\theta}^{n},\theta_{0})$
and $\mathcal{R}_{3}(\tilde{\theta}^{n},\theta_{0})$,
the results hold only asymptotically (see \cite{Khan-AoS-1973},
for asymptotic minimax efficiency of the estimator of the integral mean
of a Gaussian sample with known variance).

\section{Proofs}

\begin{pf*}{Proof of Proposition \protect\ref{Pr-ConsistencyoftheMLE}}
Under \textup{{A1}}, Kolmogo\-rov's \textsf{SLLN} implies that
$\mathbb{P}_{0}$-a.s.
$\frac{1}{n}\sum_{i=1}^{n}\ln q(Y_{i}; \theta_{j})\rightarrow\mathbb
{E}_{0}\ln q(Y;\theta_{j})$,
and for $\mathbb{P}_{0}$-a.s. any sequence of
realizations, $\hat{\theta}^{n}$ converges to $\theta_{0}$. Measurability
follows from the fact that the following set belongs to $\mathcal
{Y}^{\otimes n}$:
\begin{eqnarray*}
&&\Biggl\{ \omega\in\Omega\Big|\sup_{\theta\in\Theta}\frac{1}{n}\sum_{i=1}^{n}\ln
q(y_{i};\theta)\leq t\Biggr\} \\
&&\quad=\bigcap_{\theta_{j}\in\Theta}\Biggl\{ \omega\in
\Omega\Big|\frac{1}{n}\sum_{i=1}^{n}\ln q(y_{i};\theta_{j})\leq t\Biggr\} .
\end{eqnarray*}
\upqed\end{pf*}

\begin{pf*}{Proof of Lemma \protect\ref{Lm-Cramerconditionandeta-Int}}
Clearly \textup{(ii)} implies \textup{{A2}} for a
certain $\eta>0$. On the other hand, suppose that \textup{{A2}}
holds; then, applying recursively H\"older inequality:
\begin{eqnarray*}
\Lambda^{(i)}(\bolds{\lambda})&\triangleq&\ln\mathbb{E}_{0}\biggl[\prod
_{j=0,\ldots,J,j\neq i}\biggl(\frac{q(Y;\theta_{i})}{q(Y;\theta_{j})}\biggr)^{\lambda
_{j}}\biggr]\\
&\leq&\sum_{j=0,\ldots,J,j\neq i}\frac{1}{J}\cdot\ln\mathbb
{E}_{0}\biggl[\biggl(\frac{q(Y;\theta_{i})}{q(Y;\theta_{j})}\biggr)^{J\cdot\lambda_{j}}\biggr]
\end{eqnarray*}
and choosing the $\lambda_{j}$'s adequately, we get \textup{(ii)}.\vadjust{\goodbreak}
\end{pf*}

\begin{pf*}{Proof of Proposition \protect\ref{Pr-convergencerateofMLE2}}
The first two results are straightforward applications of Cram\'er's
Theorem in $\mathbb{R}^{d}$ (see, e.g., \cite{Dembo-Zeitouni-BOOK-1998},
Corollary 6.1.6, page 253). Indeed, it is known that the lower bound
holds without any supplementary assumption, while the upper bound
requires a Cram\'er condition $\mathbf{0}\in\operatorname{int }(\mathcal
{D}_{\Lambda^{(i)}})$;
indeed, from Lemma~\ref{Lm-Cramerconditionandeta-Int},
this is equivalent to Assumption~\textup{{A2}}. Then,
a full \textsf{LDP} holds:
\begin{eqnarray*}
&&\liminf_{n\rightarrow\infty}\frac{1}{n}\ln\mathbb{P}_{0}(\hat{\theta
}^{n}=\theta_{i}) \\[-2pt]
&&\quad \geq-\inf_{\mathbf{y}\in\operatorname{int }\mathbb
{R}_{+}^{J}}\sup_{\bolds{\lambda}\in\mathbb{R}^{J}}\bigl\{ \langle\mathbf
{y},\bolds{\lambda}\rangle-\Lambda^{(i)}(\bolds{\lambda})\bigr\} ,\\[-2pt]
&&\limsup_{n\rightarrow\infty}\frac{1}{n}\ln\mathbb{P}_{0}(\hat{\theta
}^{n}=\theta_{i})  \\[-2pt]
&&\quad\leq-\inf_{\mathbf{y}\in\mathbb{R}_{+}^{J}}\sup
_{\bolds{\lambda}\in\mathbb{R}^{J}}\bigl\{ \langle\mathbf{y},\bolds
{\lambda}\rangle-\Lambda^{(i)}(\bolds{\lambda})\bigr\} .
\end{eqnarray*}

In order to prove the final result, we have to show that $\mathbb{R}_{+}^{J}$
is a $\Lambda^{(i),\ast}$-\textit{continuity set}, that
is,\break $\inf_{\mathbf{y}\in\operatorname{int }\mathbb{R}_{+}^{J}}\Lambda
^{(i),\ast}(\mathbf{y})=\inf_{\mathbf{y}\in\mathbb{R}_{+}^{J}}\Lambda
^{(i),\ast}(\mathbf{y})$.
It is enough to apply part (ii) in Lemma on page 903 of \cite{Ney-AoP-1984}.
\end{pf*}

\begin{pf*}{Proof of Proposition \protect\ref{Pr-convergencerateofMLE}}
First of all, we note that $\mathbb{P}_{0}(\hat{\theta}^{n}\neq\theta
_{0})=\mathbb{P}_{0}(\sum_{k=1}^{n}\mathbf{X}_{k}\in\operatorname{int
}(\mathbb{R}_{+}^{J})^{c})$.
Therefore, we can apply large deviations principles, with the candidate
rate function $\Lambda^{\ast}(\mathbf{y})$; this is a
strictly convex function on $\operatorname{int }\mathcal{D}_{\Lambda^{\ast}}$
globally minimized at
\[
\mathbf{y}^{\prime}=\bigl[ \mathbb{E}_{0}\bigl(\ln q(Y;\theta_{0})-\ln q(Y;\theta
_{j})\bigr)\bigr]_{j=1,\ldots,J}.
\]
By Assumption \textup{{A1}}, $\mathbf{y}^{\prime}$ is finite
and belongs to $\operatorname{int }\mathbb{R}_{+}^{J}$. From the strict
convexity of the level sets of $\Lambda^{\ast}(\mathbf{y})$,
the set $\arg\inf_{\mathbf{y}\in\operatorname{int }(\mathbb
{R}_{+}^{J})^{c}}\Lambda^{\ast}(\mathbf{y})$
has at most finite cardinality $H$. Moreover, since large deviations
theory allows us to ignore the part of $\operatorname{int }(\mathbb{R}_{+}^{J})^{c}$
where $\Lambda^{\ast}(\mathbf{y})\geq\varepsilon+\inf_{\mathbf{y}\in
\operatorname{int }(\mathbb{R}_{+}^{J})^{c}}\Lambda^{\ast}(\mathbf{y})$,
we can replace $(\mathbb{R}_{+}^{J})^{c}$ with a collection
of $H$ disjoint sets, say $\Gamma_{h}$, $h=1,\ldots,H$, each of them
containing in its interior one and only one of the points of $\arg\inf
_{\mathbf{y}\in\operatorname{int }(\mathbb{R}_{+}^{J})^{c}}\Lambda^{\ast
}(\mathbf{y})$
(see \cite{Iltis-JTP-1995}, page 508):
\begin{eqnarray}\label{Eq-setrepresentationasGammah}
&&\mathbb{P}_{0}\Biggl(\sum_{k=1}^{n}\mathbf{X}_{k}\in\operatorname{int }(\mathbb
{R}_{+}^{J})^{c}\Biggr) \nonumber\\[-2pt]
 &&\quad=\bigl(1+o(1)\bigr)\cdot\mathbb{P}_{0}\Biggl(\sum_{k=1}^{n}\mathbf
{X}_{k}\in\operatorname{int }\bigcup_{h=1}^{H}\Gamma_{h}\Biggr)\\[-2pt]
&&\quad =\bigl(1+o(1)\bigr)\cdot\sum_{h=1}^{H}\mathbb{P}_{0}\Biggl(\sum_{k=1}^{n}\mathbf
{X}_{k}\in\operatorname{int }\Gamma_{h}\Biggr).\nonumber
\end{eqnarray}
As before, the bounds derive from Cram\'er's Theorem in $\mathbb{R}^{d}$.
Noting that the contribution of any $\Gamma_{h}$ is the same and
recalling (\ref{Eq-setrepresentationasGammah}), we get the results.\vadjust{\goodbreak}
\end{pf*}

\begin{pf*}{Proof of Proposition \protect\ref{Pr-NeyconvergencerateofMLE}}
The assumptions of the theorem on page 904 of \cite{Ney-AoP-1984}
are easily verified. This shows that a unique dominating point~$\mathbf
{y}^{(i)}$
exists and implies, through Proposition on page~161 of~\cite{Ney-AoP-1983}
(according to the ``Remarks on the hypotheses'' in \cite{Ney-AoP-1984},
page 905, the ``lattice'' conditions are not necessary), that the
stated bracketing of $\mathbb{P}_{0}(\hat{\theta}^{n}=\theta_{i})$
holds.
\end{pf*}

\begin{pf*}{Proof of Proposition \protect\ref{Pr-ExactAsymptoticsinR}}
$\!\!\!\!\!$Under Assumptions~\textup{{A1}}, \textup{{A2}}, \textup{A3} and \textup{{A4}},
according to Proposition~\ref{Pr-convergencerateofMLE2}\textup{{(iii)}}
we have $\mathbb{P}_{0}\{ Q_{n}(\theta_{1})\geq Q_{n}(\theta_{0})\}
=\mathbb{P}_{0}\{ Q_{n}(\theta_{1})> Q_{n}(\theta_{0})\} \cdot(1+o(1))$
and we can study the behavior of
\begin{eqnarray*}
\mathbb{P}_{0}(\hat{\theta}^{n}\neq\theta_{0}) & =&\mathbb{P}_{0}(\hat
{\theta}^{n}=\theta_{1})=\mathbb{P}_{0}\{ Q_{n}(\theta_{1})\geq
Q_{n}(\theta_{0})\} \\
& =&\mathbb{P}_{0}\{ Q_{n}(\theta_{1})-Q_{n}(\theta_{0})\in[0,+\infty
)\} .
\end{eqnarray*}
Assumption \textup{{A8}} implies that the conditions of Theorem 3.7.4
in \cite{Dembo-Zeitouni-BOOK-1998} (page 110) are verified, in particular
the existence of a positive $\mu\in\operatorname{int }(\mathcal{D}_{\Lambda^{(1)}})$
solution to the equation $0=(\Lambda^{(1)})^{\prime}(\mu)$.
From Lemma 2.2.5(c) in~\cite{Dembo-Zeitouni-BOOK-1998}, this implies
$\Lambda^{(1)}(\mu)=-\Lambda^{(1),\ast}(0)$,
and the result follows.
\end{pf*}\vspace*{1pt}

\begin{pf*}{Proof of Theorem \protect\ref{Th-SaddlepointApproximation}} We
note that the function $\kappa(\cdot)$ in \cite{Jing-Robinson-AoS-1994}
(page 1117) is given by\vspace*{1pt}
\begin{eqnarray*}
\kappa(\mathbf{u}) & =&\ln\mathbb{E}_{0}\exp\bigl[\mathbf{u}\cdot\bigl(\mathbf
{X}^{(i)}-\mathbb{E}_{0}\mathbf{X}^{(i)}\bigr)\bigr]\\[1pt]
& =&\ln\mathbb{E}_{0}\exp\bigl[\mathbf{u}\cdot\mathbf{X}^{(i)}\bigr]-\mathbf
{u}\cdot\mathbb{E}_{0}\mathbf{X}^{(i)}\\
& =&\Lambda^{(i)}(\mathbf{u})-\mathbf{u}\cdot\mathbb{E}_{0}\mathbf{X}^{(i)}.\vspace*{1pt}
\end{eqnarray*}
Therefore, we write the mean $\mathbf{m}(\mathbf{u})$
and covariance matrix $\mathbf{V}(\mathbf{u})$ as\vspace*{1pt}
\begin{eqnarray*}
\mathbf{m}(\mathbf{u}) & =&\kappa^{\prime}(\mathbf{u})=\frac{\partial
\kappa(\mathbf{u})}{\partial\mathbf{u}}=\frac{\partial\Lambda
^{(i)}(\mathbf{u})}{\partial\mathbf{u}}-\mathbb{E}_{0}\mathbf
{X}^{(i)},\\[1pt]
\mathbf{V}(\mathbf{u}) & =&\kappa^{\prime\prime}(\mathbf{u})=\frac
{\partial^{2}\kappa(\mathbf{u})}{\partial\mathbf{u}^{2}}=\frac{\partial
^{2}\Lambda^{(i)}(\mathbf{u})}{\partial\mathbf{u}^{2}}.\vspace*{1pt}
\end{eqnarray*}
From (\ref{Eq-ProbabilityforthetaiinLDForm}), we have\vspace*{1pt}
\begin{eqnarray*}
&&\hspace*{-3pt}\mathbb{P}_{0}(\hat{\theta}^{n}=\theta_{i})\\[1pt]
&&\hspace*{-3pt}\quad =\mathbb{P}_{0}\Biggl(\sum
_{k=1}^{n}\mathbf{X}_{k}^{(i)}\in\operatorname{int }(\mathbb{R}_{+}^{J})\Biggr)\\[1pt]
&&\hspace*{-3pt}\quad =\mathbb{P}_{0}\Biggl\{ \frac{1}{n}\cdot\sum_{k=1}^{n}\bigl(\mathbf
{X}_{k}^{(i)}-\mathbb{E}_{0}\mathbf{X}^{(i)}\bigr)\in\operatorname{int }(\mathbb
{R}_{+}^{J})\ominus\mathbb{E}_{0}\mathbf{X}^{(i)}\Biggr\} .
\end{eqnarray*}

Now we verify Assumptions (S.1)--(S.4) of \cite
{Jing-Robinson-AoS-1994}. Assumption (S.1) is implied by \textup{
{A2}}. Assumptions (S.2) and (S.3) hold since the random
vectors are i.i.d. and nontrivial.\vadjust{\goodbreak} At last, (S.4) is implied by \textsf
{\textup{A5}} (see, e.g.,~\cite{Robinson-Hoglund-Holst-Quine-AoP-1990},
page 735). Since $\mathbb{E}_{0}\mathbf{X}^{(i)}$ is strictly
negative by \textup{{A1}}, $\operatorname{int }\mathbb
{R}_{+}^{J}\ominus\mathbb{E}_{0}\mathbf{X}^{(i)}$
does not contain $\mathbf{0}$ and, according to Theorem 1 in \cite
{Jing-Robinson-AoS-1994}
(page 1118), the result of the theorem follows.
\end{pf*}

\begin{pf*}{Proof of Proposition \protect\ref{Pr-Bahadurlowerbound}} First
of all, we\break prove~(\ref{Eq-Bahadurbound-KLdivergence}). We suppose
that
\[
\int\ln\frac{f_{Y}(y;\theta_{1})}{f_{Y}(y;\theta_{0})}f_{Y}(y;\theta
_{1})\mu(\mathrm{d}y)<\infty;
\]
otherwise the inequality is trivial. Then, for any $\theta_{1}\in\Theta
\setminus\{ \theta_{0}\} $,
we apply Lemma 3.4.7 in \cite{Dembo-Zeitouni-BOOK-1998} (page~94) with
$\alpha_{n}=\mathbb{P}_{\theta_{1}}\{ \tilde{\theta}^{n}\neq\theta
_{1}\} $
and $\beta_{n}=\mathbb{P}_{\theta_{0}}\{ \tilde{\theta}^{n}\neq\theta
_{0}\} $;\break
since~$\tilde{\theta}^{n}$ is strongly consistent, $\alpha_{n}$ is
ultimately less than any $\varepsilon>0$ and the bound holds.

The second part can be proved as follows. Define the sets
\begin{eqnarray*}
A_{n}(j) & = & \{ \omega\dvtx \tilde{\theta}^{n}=\theta_{j}\}, \\[-2pt]
B_{n}(j) & = & \biggl\{ \omega\dvtx \frac{1}{n}\ln\biggl(\frac{f_{Y}(Y;\theta
_{j})}{f_{Y}(Y;\theta_{0})}\biggr)\\[-2pt]
&&\hspace*{6pt}\le\mathbb{E}_{\mathbb{\theta}_{j}}\ln\biggl(\frac
{f_{Y}(Y;\theta_{j})}{f_{Y}(Y;\theta_{0})}\biggr)+\varepsilon\biggr\}.
\end{eqnarray*}
Therefore, we have
\begin{eqnarray*}
& & \mathbb{P}_{\theta_{0}}\{ \tilde{\theta}^{n}\neq\theta_{0}\}\\[-1pt]
&&\quad=\mathbb{E}_{\theta_{0}}\mathsf{1}\{ \tilde{\theta}^{n}\neq\theta_{0}\}
\\[-1pt]
&&\quad=\mathbb{E}_{\theta_{j}}\frac{f_{Y}(Y;\theta_{0})}{f_{Y}(Y;\theta
_{j})}\mathsf{1}\{ \tilde{\theta}^{n}\neq\theta_{0}\} \\[-1pt]
&&\quad \ge \mathbb{E}_{\theta_{j}}\frac{f_{Y}(Y;\theta
_{0})}{f_{Y}(Y;\theta_{j})}\mathsf{1}\{ A_{n}(j)\} \\[-1pt]
&&\quad \ge \mathbb{E}_{\theta_{j}}\mathsf{1}\{ A_{n}(j)\} \mathsf{1}\{
B_{n}(j)\}\\[-1pt]
&&\qquad{} \cdot\exp\biggl\{ -n\cdot\biggl[\mathbb{E}_{\mathbb{\theta}_{j}}\ln\biggl(\frac
{f_{Y}(Y;\theta_{j})}{f_{Y}(Y;\theta_{0})}\biggr)+\varepsilon\biggr]\biggr\} \\[-1pt]
&&\quad \ge [1-\mathbb{P}_{\theta_{j}}\{ A_{n}^{c}(j)\} -\mathbb{P}_{\theta
_{j}}\{ B_{n}^{c}(j)\} ]\\[-1pt]
&&\qquad{}\cdot\exp\biggl\{ -n\cdot\biggl[\mathbb{E}_{\mathbb{\theta
}_{j}}\ln\biggl(\frac{f_{Y}(Y;\theta_{j})}{f_{Y}(Y;\theta_{0})}\biggr)+\varepsilon
\biggr]\biggr\} \\[-1pt]
&&\quad \ge [1-\mathbb{P}_{\theta_{j}}\{ \tilde{\theta}^{n}\neq\theta
_{j}\} -\mathbb{P}_{\theta_{j}}\{ B_{n}^{c}(j)\} ]\\[-1pt]
&&\qquad{}\cdot\exp\biggl\{ -n\cdot
\biggl[\mathbb{E}_{\mathbb{\theta}_{j}}\ln\biggl(\frac{f_{Y}(Y;\theta
_{j})}{f_{Y}(Y;\theta_{0})}\biggr)+\varepsilon\biggr]\biggr\} .
\end{eqnarray*}
This implies:
\begin{eqnarray*}
&&\liminf_{n\rightarrow\infty}\frac{1}{n}\ln\mathbb{P}_{\theta_{0}}\{
\tilde{\theta}^{n}\neq\theta_{0}\} \\[-2pt]
&&\quad \ge -\mathbb{E}_{\mathbb{\theta
}_{j}}\ln\biggl(\frac{f_{Y}(Y;\theta_{j})}{f_{Y}(Y;\theta_{0})}\biggr)-\varepsilon\\[-2pt]
& &\qquad{} +\liminf_{n\rightarrow\infty}\frac{1}{n}\ln[1-\mathbb{P}_{\theta
_{j}}\{ \tilde{\theta}^{n}\neq\theta_{j}\} -\mathbb{P}_{\theta_{j}}\{
B_{n}^{c}(j)\} ].\vadjust{\goodbreak}
\end{eqnarray*}
Now, since $\lim_{n\rightarrow\infty}\mathbb{P}_{\theta_{j}}\{
B_{n}^{c}(j)\} =0$
and\break $\limsup_{n\rightarrow\infty}\mathbb{P}_{\theta_{j}}\{ \tilde{\theta
}^{n}\neq\theta_{j}\} <1$,
the third term in the right-hand side goes to zero; since $\varepsilon$
is arbitrary, the result follows.
\end{pf*}

\begin{pf*}{Proof of Proposition \protect\ref{Pr-Chapman-Robbinslowerbound}}
From the Neyman--Pearson Lemma, we have
\begin{eqnarray*}
&&\sup_{\theta_{0}\in\Theta}\mathbb{P}_{\theta_{0}}(\tilde{\theta}^{n}\neq
\theta_{0}) \\
 &&\quad\geq\max\{ \mathbb{P}_{\theta_{0}}(\tilde{\theta}^{n}\neq
\theta_{0}),\mathbb{P}_{\theta_{1}}(\tilde{\theta}^{n}\neq\theta_{1})\}
\\
&&\quad\geq\frac{1}{2}\cdot\{ \mathbb{P}_{\theta_{0}}(\tilde{\theta}^{n}\neq
\theta_{0})+\mathbb{P}_{\theta_{1}}(\tilde{\theta}^{n}\neq\theta_{1})\}
\\
&&\quad \geq\frac{1}{2}\cdot\biggl\{ \mathbb{P}_{\mathbb{\theta}_{0}}\biggl(\frac
{\mathsf{L}_{n}(\theta_{0})}{\mathsf{L}_{n}(\theta_{1})}<1\biggr)+\mathbb
{P}_{\mathbb{\theta}_{1}}\biggl(\frac{\mathsf{L}_{n}(\theta_{0})}{\mathsf
{L}_{n}(\theta_{1})}\geq1\biggr)\biggr\}
\end{eqnarray*}
for an arbitrary couple of different alternatives $\theta_{0}$ and
$\theta_{1}$ in $\Theta$. Then we can use Chernoff's Bound (\cite
{Dembo-Zeitouni-BOOK-1998},
page 93); the final expression derives from the equality $\Lambda
^{*}(0)=-\inf_{\lambda\in\mathbb{R}}\Lambda(\lambda)$.
\end{pf*}

\begin{pf*}{Proof of Proposition \protect\ref{Pr-AdmissibilityandMinimaxEfficiencyofMLE}}
In order to prove that the \textsf{MLE} is admissible and minimax
we use the Bayesian method. Using the prior densities given by $\pi
(\theta_{k})=(J+1)^{-1}$,
the Bayes estimator relative to zero--one loss $\check{\theta}^{n}$
coincides with the \textsf{MLE} $\hat{\theta}^{n}$. Therefore, respectively
from Lemma 2.10 and Proposition~6.3 in \cite{Robert-BOOK-1994}, $\hat
{\theta}^{n}$
is minimax and admissible
. The fact that the \textsf{MLE} minimizes the average
probability of error derives from Proposition~\ref{Pr-MinimizationoftheBayesRisk}.
\end{pf*}

\begin{pf*}{Proof of Proposition \protect\ref{Pr-Proofofnon-IR-optimality}}
\textup{(i)} In order to prove the first statement, we apply Lemma
2.4 in \cite{Kester-Kallenberg-AoS-1986} (page~653). Clearly $\mathcal{P}$
is closed in total variation, since it is finite, and is not exponentially
convex; indeed, under Assumption \textup{{A7}}, there exist
$\theta_{1},\theta_{2}\in\Theta$ and $\alpha\in[0,1]$,
such that the probability measure~$\mathbb{P}_{\theta(\alpha)}$
defined as
\[
\mathbb{P}_{\theta(\alpha)}(\mathrm{d}x)=\frac{(f_{\theta
_{1}}(x))^{\alpha}\cdot(f_{\theta_{2}}(x))^{1-\alpha}}{\int(f_{\theta
_{1}}(x))^{\alpha}\cdot(f_{\theta_{2}}(x))^{1-\alpha}\cdot\mu(\mathrm
{d}x)}\mu(\mathrm{d}x)
\]
does not belong to $\mathcal{P}$. Therefore, from Lemma 2.4\textup{(iii)}
in \cite{Kester-Kallenberg-AoS-1986}, there exist $\theta_{1}^{\prime
},\theta_{2}^{\prime}\in\Theta$
such that Equation~(2.12) in \cite{Kester-Kallenberg-AoS-1986} holds
and, as a consequence of Lemma~2.4\textup{(i)} in \cite
{Kester-Kallenberg-AoS-1986},
the \textsf{MLE} fails to be an inaccuracy rate optimal estimator
at least at one of the points $\theta_{1}^{\prime},\theta_{2}^{\prime}$.
This means that, say for $\theta_{1}^{\prime}$:
\begin{eqnarray*}
&&\liminf_{n\rightarrow\infty}\frac{1}{n}\ln\mathbb{P}_{\theta_{1}^{\prime
}}\{ |\hat{\theta}^{n}-\theta_{1}^{\prime}|>\varepsilon\} \\
&&\quad>\sup_{\theta
\in\Theta,|\theta-\theta_{1}^{\prime}|>\varepsilon}\mathbb{E}_{\mathbb
{\theta}}\ln\biggl(\frac{f_{Y}(Y;\theta_{1}^{\prime})}{f_{Y}(Y;\theta)}\biggr),
\end{eqnarray*}
and this implies that the Chapman--Robbins bound is not attained at
$\theta_{1}^{\prime}$.\vspace*{-6pt}\vadjust{\goodbreak}
\begin{longlist}[(iii)]
\item[(ii)] The second statement follows easily from the results
of \cite{Kanaya-Han-IEEETIT-1995} (Theorem 2) on $\lim_{n\rightarrow
\infty}\frac{1}{n}\ln r_{1}(\tilde{\theta}^{n},\pi)$,
using Equation (\ref{Eq-AsymptoticEquivalenceofBayesRiskandMinimaxProbError}).
Indeed, the \textsf{MLE} attains the lower bound (\ref{Eq-ChernoffsLowerBound})
and is therefore asymptotically minimax efficient.

\item[(iii)] If the estimator is asymptotically $\mathrm{CR}$-efficient
w.r.t. $\mathcal{R}_{1}$ at $\theta_{0}$, this means that
at $\theta_{0}$ it is more efficient than the \textsf{MLE} and therefore
it has to be less efficient elsewhere (since from Proposition~\ref{Pr-AdmissibilityandMinimaxEfficiencyofMLE}
the \textsf{MLE} minimizes the probability of error). Therefore, it
cannot be minimax $\mathrm{CR}$-efficient.\qed
\end{longlist}
\noqed\end{pf*}

\begin{pf*}{Proof of Proposition \protect\ref{Pr-Superefficiency}} For \textup
{(i)} it is enough to follow the proof of Proposition~\ref{Pr-Bahadurlowerbound} and to reason by contradiction, while \textup
{(ii)} is
simply another way of stating Proposition
\ref{Pr-AdmissibilityandMinimaxEfficiencyofMLE}.
\end{pf*}

\section*{Acknowledgments}
The authors would like to thank Lucien Birg\'e,
Mehmet Caner, Jean-Pierre Florens, Christian Gou\-ri\'eroux, Christian
Hess, Marc Hoffmann, Pierre Jacob, S\o ren Johansen, Rasul A. Khan,
Oliver B. Linton, Christian P. Robert, Keunkwan Ryu, Igor Vajda and
the participants to seminars at Universit\'e Paris 9 Dauphine, CREST
and Institut Henri Poin\-car\'e, to ESEM 2001 in Lausanne, to XXXIV\`emes
Journ\'ees de Statistique 2002 in Bruxelles, to BS/\allowbreak IMSC 2004 in Barcelona,
and to ESEM 2004 in Madrid. All the remaining errors are our responsibility.


\end{document}